\documentclass[lettersize,journal]{IEEEtran}

\usepackage{amsmath,amsfonts}
\usepackage{algorithmic}
\usepackage{algorithm}
\usepackage{array}
\usepackage[caption=false,font=normalsize,labelfont=sf,textfont=sf]{subfig}
\usepackage{textcomp}
\usepackage{stfloats}
\usepackage{verbatim}
\usepackage{cite}
\usepackage[english]{babel}
\usepackage{blindtext}

\usepackage{hyperref}
\usepackage{url}
\usepackage{cleveref}
\usepackage{makecell}

\usepackage{graphicx}
\usepackage{caption}

\graphicspath{ {./sections/figures/} }
\usepackage{epstopdf}
\DeclareGraphicsExtensions{.png,.pdf}

\usepackage{pdfpages}

\usepackage{booktabs}
\usepackage{tablefootnote}
\usepackage{multirow}
\usepackage{longtable}
\usepackage{caption}

\usepackage{soulutf8}
\usepackage{xcolor}

\soulregister{\cite}{7}
\soulregister{\ref}{7}
\soulregister{\pageref}{7}
\soulregister{\mbR}{1}

\newcommand{\mbR}[1]{{\color{black}#1}}

\begin{document}
\title{adF: a Novel System for Measuring Web Fingerprinting through Ads}


\author{Miguel A Bermejo-Agueda,
\thanks{M. Bermejo-Agueda is with the Telematic Engineering Department, Universidad Carlos III of Madrid, 28911 Leganés, Spain, email: mibermej@pa.uc3m.es.}
\and
Patricia Callejo,
\thanks{P. Callejo is with the Telematic Engineering Department, Universidad Carlos III of Madrid, 28911 Leganés, Spain, email: pcallejo@it.uc3m.es.}
\and
Rubén Cuevas,
\thanks{R. Cuevas is with the Telematic Engineering Department, Universidad Carlos III of Madrid, 28911 Leganés, Spain, email: rcuevas@it.uc3m.es.}
\and
Ángel Cuevas,
\thanks{A. Cuevas is with the Telematic Engineering Department, Universidad Carlos III of Madrid, 28911 Leganés, Spain, email: acrumin@it.uc3m.es.}
\and
Álvaro Mayol
\thanks{Á. Mayol is with Taptap Digital, 28020 Madrid, Spain, email: alvaro.mayol@taptapdigital.com.}}

\markboth{2024 IEEE TETC on Cyber Security and Resilience (CSR)}%
{Shell \MakeLowercase{\textit{Bermejo-Agueda et al.}}: A Sample Article Using IEEEtran.cls for IEEE Journals}


\maketitle

\begin{abstract}
This paper introduces \emph{adF}, a novel system for analyzing the vulnerability of different devices, operating systems, and browsers to web fingerprinting. \emph{adF} performs its measurements from code inserted in ads. We use our system in several ad campaigns that delivered \mbR{5.40} million ad impressions. The collected data allow us to assess the vulnerability of current desktop and mobile devices to web fingerprinting. We estimate that \mbR{66\%} of desktop and \mbR{40\%} of mobile devices can be uniquely fingerprinted with our web fingerprinting system. However, web fingerprinting resilience varies significantly across browsers and device types, with Chrome on desktops being the most vulnerable configuration.

To counter web fingerprinting, we propose \emph{ShieldF}, a straightforward solution that blocks browsers from reporting attributes with the most significant discrimination power, as identified through our dataset analysis. Our experiments reveal that \emph{ShieldF} outperforms all anti-fingerprinting solutions proposed by major browsers (Chrome, Safari and Firefox), providing a resilience boost of up to \mbR{62\%} for some device configurations. \emph{ShieldF} is accessible as an add-on for any Chromium-based browser. Moreover, it is readily adoptable by browser and mobile app developers. Its widespread use would lead to a marked enhancement in browsers' and mobile apps' defense against web fingerprinting.


\end{abstract}


\section{Introduction}

\IEEEPARstart{D}{igital} advertising is arguably the primary source of funding for the current Internet. Some of the most important tech companies, such as Google or Meta (previously Facebook), declare that over \mbR{77.5\% and \mbR{97.5}\% (2023 data) of their profits come from digital advertising} \cite{advertising_revenue_2023_google, advertising_revenue_2023_meta}, respectively.

The appeal of online advertising for advertisers, compared to traditional forms of publicity (e.g., TV, radio, or newspapers), lies in its ability to deliver personalized ads. Implementing personalized advertising is based on techniques that identify unique users online. These techniques are used for tracking, profiling users, displaying targeted or re-targeted ads, implementing attribution models, etc. The most widely used techniques for this purpose are third-party cookies, advertising IDs, and fingerprinting \cite{bujlow2017survey}.

Third-party cookies and advertising IDs, when properly implemented with consent acquisition (not always the case), comply with data protection regulations. Fingerprinting, by definition, is an intrusive practice and represents a severe threat to end-users' privacy. The most common type of fingerprinting involves embedding a script in a website. When a user visits such a website, the script collects attributes from the browser, OS, and device rendering the web-page. The concatenation of these attributes creates a device fingerprint. If the number of attributes is large enough, the fingerprint can make the device unique even among hundreds of thousands or millions of devices. This practice is known as \emph{web fingerprinting}. The privacy risks associated with web fingerprinting have prompted extensive research in this area \cite{mayer2009any, eckersley2010, acar2014web, laperdrix2016beauty, gomezBoix2018hiding, cao2017cross}.
A body of work proposes countermeasure techniques to prevent fingerprinting \cite{laperdrix2015mitigating, Al_FannahNasserMohammed2020Tltl, azad2020taming, canvasbloquer_firefox, wu2019rendered, laperdrix2016beauty, iqbal2021fingerprinting, privacybadger, nikiforakis2015privaricator, laperdrix2017fprandom, noscript}. Browser developers (e.g., Apple, Mozilla, or Google) have also proposed anti-fingerprinting measures in later releases of their browsers~\cite{cameron_2018_appleagainstfp, enhanced_tracking_protection_mozillawiki, user_agent_reduction_privacysandbox}. However, a rapid web technology development has made some existing proposals obsolete and requires a continuous effort from browser developers to update their anti-fingerprinting solutions.

In this rapidly evolving context, proper auditing methodologies are crucial for measuring the risk of web fingerprinting to users and assessing vulnerabilities in different web browsers. The research community has proposed measurement solutions in this direction \mbR{\cite{eckersley2010, laperdrix2016beauty, boda2011user, cao2017cross}}. Essentially, these solutions involve deploying a script on a web-page to compute the device fingerprint when a device connects to the page. This methodology assesses whether the computed fingerprint is unique within the database of previously computed fingerprints.
This approach has been very useful in improving our understanding of web fingerprinting. However, it presents some limitations. On the one hand, if the auditing script is deployed on a website meant for measuring fingerprinting \cite{eckersley2010, laperdrix2016beauty}, it requires the voluntary participation of users who are willing to assess their fingerprint. Sustaining a continuous flow of visitors to these types of website over time poses a challenge. On the other hand, running the script on a regular website \cite{gomezBoix2018hiding, andriamilanto2021largescale, li2020wtmbf} introduces biases in the type of devices/users visiting that single website. Moreover, the ability to conduct such experiments depends on the willingness of the website owner to participate. 
Furthermore, the coverage of this technique is limited to web-browsers, leaving aside other very common venues such as mobile apps that operate on top of WebViews, essentially utilizing web technologies.

In this paper, we propose a novel measurement technology to study web fingerprinting. Our methodology utilizes ads, which are HTML elements included on web-pages and within mobile apps. We embed a script to compute the device fingerprint within these ads, overcoming limitations of existing solutions: 1) We run ad campaigns that perform measurements transparently to users, ensuring longitudinal analyses without relying on volunteers; 2) Our methodology allows for the study of web fingerprinting within the context of mobile apps; 3) Our data sample is not limited to measurements from a single vantage point (i.e., web-page); 4) Researchers have full control over our technique, deciding when to run the campaigns; 5) By harnessing the targeting capabilities of online advertising, our methodology enables the definition of specific device configurations (based on device type, OS, and browser) to be measured for the first time.

We implement our methodology in a system we refer to as \emph{adF} (as discussed in Section \ref{sec:system}) and conducted real ad campaigns that collected more than \mbR{4.04M} fingerprint samples in three independent periods of time over the course of a year and a half (as detailed in Section \ref{subsec:ad_campaigns}). First, these real experiments demonstrate that our methodology achieves an accuracy exceeding \mbR{80\%}, confirming its validity for the study of web fingerprinting.
Second, the analysis of the collected fingerprint samples allows to assess the vulnerability of different device configurations to web fingerprinting. We define a device configuration as the combination of device type (desktop vs. mobile), an operating system (e.g., Windows, macOS, Android, or iOS), and a browser (e.g., Mozilla Firefox, Chrome, or Safari). In the case of mobile apps, we only distinguish between operating systems, Android vs. iOS (see Section \ref{sec:results}).

Our analysis reveals that \mbR{66\%} of desktop devices can be uniquely fingerprinted from ads rendered by browsers. In the case of mobile devices, \mbR{40\% (35\%)} can be fingerprinted with ads delivered to browsers (mobile apps). Moreover, we observe varying degrees of vulnerability to fingerprinting among different device configurations. For instance, in mobile devices, iOS offers significantly better protection against web fingerprinting compared to Android. In the case of desktop devices, over \mbR{78\%} of devices running Chrome can be fingerprinted, regardless of the operating system. This percentage drops to around \mbR{5.5\%} for \texttt{\{Safari, macOS\}} and stands at \mbR{22\%} for \texttt{\{Firefox, Linux\}}.
Finally, a thorough analysis of all attributes contributing to web fingerprinting reveals that cardinality (number of distinct values an attribute can take) and entropy (frequency of appearance of the different values) are the two crucial parameters that define the discrimination power of attributes. The device configurations most vulnerable to fingerprinting are those associated with a higher number of attributes with significant discrimination power (as discussed in Section \ref{sec:atributes_analysis}).
Building on this insight, we propose \emph{ShieldF} a straightforward approach, easily adoptable by browser developers, to counter web fingerprinting. ShieldF is a browser extension compatible with all Chromium-based browsers. It prevents the reporting of attributes with high discrimination power to third-parties while preserving the user's browsing experience. Our experiments show that \emph{ShieldF} outperforms most popular countermeasure defenses in the literature, and the anti-fingerprinting solutions of all major browser developers (Section \ref{sec:solution}), enhancing the resilience to web fingerprinting across all considered device configurations by up to \mbR{62\%}.
  
\section{Background}
\label{sec:background}

\subsection{Identifying individual users}

Advertising companies use different techniques to identify unique users and target them based on individual preferences such as user demographics, location, and interests. Although they can generally only identify devices rather than users (as in the open advertising ecosystem of websites and mobile apps, there is no registration process), they assume that a single user uses a device. According to \cite{bujlow2017survey}, the most common techniques used to identify unique users are: third-party cookies, advertising IDs, and web fingerprinting techniques. 

\subsubsection{Third-party cookies}
A cookie is a small file placed by third-parties on the browser that serves as a unique identifier for a user. Whenever the browser visits a website where the third-party is present, the third-party receives its cookie and knows the user visited the current website. Using this technique, the third-party company can (partially) reconstruct the users' web browsing history in the specific browser where the cookie is placed. In addition, if the third-party can classify the visited websites into categories, it can infer the person's interests. Third-party cookies are also used for other purposes, such as re-targeting, which allow companies to remind users of their products or services through ads, emails, social media, or other methods.

\subsubsection{Advertising ID}
The advertising ID (Ad-ID) is a unique user identifier from the UUID standard assigned to an electronic device (smartphone, tablet, PC) and is intended for advertising purposes. It allows advertisers to track users' advertising activity to personalize offers through browsers and mobile apps. The Ad-ID is embedded in the ad requests received by ad-tech stakeholders, which use the Ad-ID to enrich the information known about such Ad-ID, e.g., demographic information or interests associated to the Ad-ID. The Ad-ID is available in all major desktop and mobile operating systems (Windows, iOS, macOS, Android, etc.) and usually receives a different name for each operating system, for instance, AAID (Android Advertising ID) in Android or IDFA (Identifier for Advertisers) in Apple's operating systems. 

\subsubsection{Web fingerprinting}
Web browsers have developed a set of APIs that allow accessing different parameters of the browser, the operating system, or the device. The original goal of these functions is to optimize the functionality of browsers and the user experience. For instance, one of these functions exposes the device's screen size such that the web-page's content can be appropriate for the specific screen size.

Web fingerprinting is implemented through a script embedded in the HTML code of the web-page. This script leverages the different functions and APIs offered by the browser to collect different attributes from the browser configuration (e.g., browser add-ons or configured languages), the OS (installed fonts), and the device (sound device or graphic card information). In conjunction, the values of all these attributes create a fingerprint of the device. If a sufficiently large number of attributes are collected, the resulting fingerprint of the device might be unique even among a large pool of devices.

Note that, contrary to the case of third-party cookies or Ad-IDs, which require implicit or explicit consent from the user, fingerprinting is implemented without any type of consent from the user and thus it is the most privacy-aggressive practice of the three. Nonetheless, web fingerprinting also has legit uses, such as pursuing fraud or credential theft, by verifying the legitimacy of users when logging into specific sites \cite{gabryel2020browser}.

\subsection{Web measurements from ads}

The online advertising industry has evolved into an intricate ecosystem for measuring user engagement, drawing on data from JavaScript scripts embedded within ads delivered to websites and mobile apps. These scripts, belonging to different third-party companies, exploit browser APIs to collect data, which is then processed and sent to third-parties' back-end servers. These scripts can be used to measure KPIs (e.g., the click-through ratio of an ad campaign), detect fraudulent practices, and more.
In this paper, we use this same measurement technology to implement our script to study web fingerprinting. We embed a script in an ad (rather than a web-page) and gather information about the browser configuration, OS, and device software and hardware while aiming to identify a user uniquely. Nonetheless, note scripts executed within ads face limitations compared to those running natively in a website's HTML code due to security measures like the Cross-Domain Policy.

\subsubsection{Web fingerprinting from ads}
One of the main uses of web fingerprinting is to reconstruct the web browsing history of a user. Tracking entities embed their fingerprinting script in as many websites as possible to gather data about the user's browsing history. It can also be used for re-targeting, i.e., users are shown ads about products they previously viewed on other websites or e-commerce sites. However, web fingerprinting may not be effective in these tasks since it is not feasible to show the same ad to a user on every page they visit. Nonetheless, web fingerprinting from ads can be very useful in building attribution models for advertisers.
An attribution model is an extensively used and very valuable tool used by advertisers to understand the performance of their ad campaigns, web-pages, and ad channels. To build this model, advertisers must reconstruct each user's path. This is a vector including all the ads shown to an individual user and the interaction of the user with them. By analyzing the paths of thousands of users, a model can be constructed to infer the most successful elements (ad creatives, ad campaigns, web-pages, mobile apps, etc) that lead to more conversions, such as purchases.

To create users' paths, it is crucial to uniquely identify each user. In this context, web fingerprinting from ads is a plausible alternative to building attribution models. Even with the imminent discontinuation of third-party cookies~\cite{blocked_chrome_3partycookie_2024}, and the significant restrictions imposed on the use of the Ad-ID~\cite{advertisingID_announcementGoogle} which limits the number of devices disclosing it\footnote{\mbR{Just 48.52\% of the 4.04M samples collected in the dataset used in this paper reported the Ad-ID.}}, it may become the only practical approach to building attribution models in the near future. 

In summary, web fingerprinting from ads also poses a severe threat to end-users privacy, while there is a clear incentive (attribution models) for advertising stakeholders to (potentially) use it. Our main goal is to assess the vulnerability of different device configurations to web fingerprinting. In the context of this paper, a device configuration is defined by the combination of three elements: device type (mobile vs. desktop), operating system (e.g., Windows or iOS), and browser (e.g., Chrome, Safari, or Firefox). In addition, we also consider the specific case of apps on mobile devices (Android and iOS).

\subsection{Popular browser' approaches to counter fingerprinting}
\label{sec:browsers_approaches}

Due to the threat that fingerprinting represents to users' privacy and the social pressure generated, browser developers have proposed different solutions to counter web fingerprinting. These techniques propose modifying, overriding, randomizing, or blocking the value reported for certain attributes. Next, we briefly describe the solutions defined by different browsers to counter web fingerprinting: 

\emph{- Brave}: While Brave is subject to web fingerprinting, it is not exposed to web fingerprinting from ads since the advertisements shown by brave are simple text notifications where code cannot be embedded. To mitigate web fingerprinting, Brave blocks several APIs (such as Canvas, Battery Status, WebGL, and Web Bluetooth, among others). These functions are natively integrated into the browser~\cite{brave_fp_protection_mode}. 

\emph{- Tor}: It is the browser proposing the most robust solution against fingerprinting. Its approach is to build the same fingerprint for all devices on the web \cite{laperdrix2020fpSurvey}. However, this approach requires blocking some essential functions, which affects the user's browsing experience \cite{khattak2016you}.

\emph{- Safari}: In 2018, Apple released a version of Safari with an anti-fingerprinting setting already built into the browser and thus requiring no user action \cite{cameron_2018_appleagainstfp}. In particular, Safari does not support specific APIs such as Web Bluetooth, Device Memory, or User Permission State (e.g., camera, geolocation, or microphone). It also removes detailed information on the user agent or on the type of graphic card used. 

\emph{- Firefox}: Mozilla Firefox offers user-configurable privacy options through fingerprint controls. 
In 2019, Firefox enabled the Enhanced Tracking Protection feature by default as part of its settings \cite{enhanced_tracking_protection_mozillawiki}. 
Today, this functionality offers opt-in fingerprint protection preventing the collection of information related to time zone, installed fonts, window size preferences, and operating system versions (among others).


\emph{- Chrome}: It is the browser with the largest market share. Google announced in August 2019 that they would introduce fingerprint-blocking features in Chrome but without providing an expected release date. It was not until early 2020 that Google revealed their plans: reduce the granularity of the information provided by the user agent, which is sent through every HTTP request by default \cite{user_agent_reduction_privacysandbox}. This functionality is still in the experimental phase. 


\emph{- Edge}: It is a web browser developed by Microsoft based on the Chromium engine. 
Then, it offers similar levels of protection to fingerprinting as Google Chrome.

\emph{- MiuiBrowser}: This Xiaomi-developed browser is a non-exclusive browser for MIUI. While we have been unable to find any officially declared anti-fingerprint technique for this browser, its advanced configuration options allow the change of the user agent by setting the value to desktop mode, iPhone, or iPad.

\vspace{2mm}
All these solutions have been designed to counter web fingerprinting by blocking access to certain APIs or functions from any code run by the web browser. This includes code run from ads. Therefore, they are also valid for the case of web fingerprinting from ads (with the exception of Brave, where web fingerprinting from ads does not apply).

Finally, in the case of mobile apps, operating systems such as iOS and Android are implementing significant changes to increase privacy and security by restricting access to data that could be used for fingerprinting \cite{yodelmobile2023}. \mbR{Based on our results}, we find that iOS and Android restrict access to information that might be used for fingerprinting from their mobile applications. 

\section{Threat Model for the \emph{\MakeLowercase{ad}F}}
\label{subsec:threatModel}

In this section, we detail the specific threat model for \emph{adF}, the system we present in Section \ref{sec:system} to evaluate the vulnerability of different device configurations to web fingerprinting.

\subsection{System model}

Our system model consists of a user who uses a (web) browser, such as Chrome, Safari, Firefox, etc., to browse the Internet. The browser runs on any supported device, including desktops, laptops, smartphones, or tablets, and can be run on any relevant operating system such as Windows, macOS, Linux, iOS, or Android. The user can also use mobile applications that operate on Android or iOS WebView.
As users access web-pages or mobile applications, they may encounter both first-party and third-party JavaScript code. The browser or app executes the code using its JavaScript engine, such as V8 in Chrome, Edge, or the Android WebView; JavaScriptCore (also known as Nitro) in Safari and iOS WebView; or SpiderMonkey in Firefox. The execution of this code enables interaction with the browser and device through various supported JavaScript APIs. For instance, it can retrieve the user-agent information, screen dimensions, audio and video settings, etc.

\subsection{Attacker model}

We consider a \emph{remote} attacker who wants to track a web user through \emph{web fingerprinting} techniques embedded in advertisements. In the threat model, the attacker is a malicious ad-tech provider or any other malicious player with access to advertising platforms capable of running ad campaigns. The attacker can add a JavaScript (JS) pixel into ads, which is widely accepted and not considered an illegal technique in ad-tech. This makes it relatively easy for the attacker to perform the attack without the need for third-party collaboration or black-hat hacking techniques to gain access to any third-party venue (e.g., a web-page). This approach significantly lowers the barriers to conducting such attacks, allowing potential reach to tens of millions of users for fingerprinting purposes.

In contrast, the attacker model defined in previous works in the literature \cite{eckersley2010, boda2011user, laperdrix2016beauty, gomezBoix2018hiding, laperdrix2021fingerprinting, andriamilanto2021largescale} requires the attacker to gain access to one or multiple web-pages to inject a fingerprinting script. This requires either the cooperation of a third party, such as the web-page owner—who may be hesitant due to data protection concerns (e.g., GDPR compliance)—or illegally gaining access to the website by black-hat hacking techniques. To reach tens of millions of users, the accessed websites should be relatively popular, which (in principle) makes it even harder to reach an agreement with their owners due to potential bad reputation implications or hack them (since it is assumed popular web-pages have better security measures in place compared to unpopular ones). Therefore, traditional threat models presented in the literature depict a much higher barrier to performing a fingerprinting attack compared to the alternative using ads discussed in this paper.

\section{\emph{\MakeLowercase{ad}F}: A system to measure web fingerprinting from ads}
\label{sec:system}

\begin{figure}[t]
\centering
\includegraphics[width=\columnwidth]{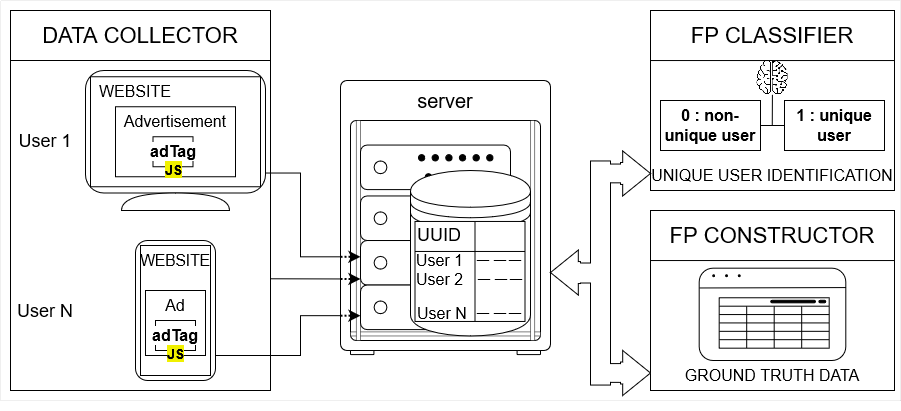}
\caption{\emph{adF} visual representation, our web fingerprinting system.} 
\label{fig:adFP_scheme}
\end{figure}

\emph{adF} is a novel system designed to assess the vulnerability of different device configurations to web fingerprinting. Figure \ref{fig:adFP_scheme} shows the scheme of the \emph{adF} system, formed by three components: i) The \emph{Data Collector}, a piece of code embedded in online ads that allows collecting a large number of attributes from the browser or the app rendering the ad; ii) The \emph{Fingerprint Constructor}, receives as input the value of the attributes collected by the Data Collector and uses encoding techniques to create a fingerprint; iii) The \emph{Fingerprint Classifier}, decides whether a fingerprint is a unique fingerprint, and thus uniquely identifies a user, or not. The final output of our system is a device fingerprint, categorized as either unique or non-unique.

\subsection{Data Collector}

To collect attributes from the device, we develop an \emph{adTag}, a JavaScript (JS) code, which can be embedded into any ad format capable of running a script. Our \emph{adTag} is executed upon the rendering process of the ad on a device. The use of adTags is a widespread practice in online advertising. Moreover, the research community has also used adTags to analyze transparency aspects of online advertising \cite{callejo2019q} and to conduct network measurements \cite{callejo2017opportunities}.

Our \emph{adTag} captures a comprehensive set of up to \mbR{191} diverse attributes from the device, OS, and browser where the ad is displayed. Specifically, it leverages JavaScript APIs within browsers to retrieve information. While some APIs necessitate permissions for access, such as the microphone, camera, or location, the majority of them are openly accessible through our \emph{adTag}.
The attributes retrieved by our \emph{adTag} via these APIs are stored on a server within our domain, where we process them and construct the fingerprint using the Fingerprint Constructor. 
Table \ref{tab:considered_attributes} in Appendix \ref{appendix:a} provides the complete list of attributes considered, along with their respective sources.

An initial analysis of the \mbR{191} considered attributes, based on tens of thousands of collected samples, reveals that some of these attributes are not suitable for creating unique fingerprints. Therefore, our Data Collector does not retrieve attributes that meet any of the following criteria: 1) the attribute reports a constant value (e.g., Bars settings: \texttt{menubar}); 2) the value of the attribute is unstable over time (e.g., Network Information: \texttt{downlink}); 3) the value is unreliable (e.g., Window \texttt{navigator.onLine}); 4) most popular browsers do not support the attribute (e.g., WebHID API). After applying this filtering, the final version of the Data Collector retrieves a total of \mbR{66} attributes from the initial \mbR{191}. These attributes are highlighted in bold in Table \ref{tab:considered_attributes}. Note that some of these \mbR{66} attributes are joined to form meta-attributes defined as a concatenation of several individual attributes. Table \ref{tab:considered_attributes} also details the mapping of attributes to meta-attributes. 

In the case of mobile apps, just a subset of \mbR{35} of these \mbR{66} attributes are considered and thus used in our analyses. These are identified in Table \ref{tab:considered_attributes} with an asterisk (*).

Ads are expected to be exposed on a device for a limited time. Usually, a few seconds \cite{exposure_time_ads}. Therefore, our adTag must collect the \mbR{66 (35)} attributes and deliver them to the backend server before the user closes the website (or the mobile app) where the ad is displayed. To this end, we use AJAX (Asynchronous JavaScript and XML), which allows the asynchronous execution of code so that the most time-consuming operations do not block the rest of the script's tasks. By doing so, we ensure that for more than \mbR{95\%} of the ads we received a response from the browser (mobile app) including the complete list of attributes. Note that we discard incomplete fingerprint samples to avoid generating fingerprints that may differ due to these uncollected attributes.

\subsection{Fingerprint Constructor}

This module aims to build a device's fingerprint using the \mbR{66 (35)} attributes collected by the Data Collector from the browser (mobile app). To this end, the Fingerprint Constructor first generates a string concatenating the value of these \mbR{66 (35)} attributes. On average, this string has a size of \mbR{657 (489)} Bytes among the samples available in our dataset (as described in Section~\ref{subsec:ad_campaigns}). We could use this string as a fingerprint of the device. However, for enhanced computational efficiency and integrity, we apply the SHA-256 cryptographic hash function. This creates a 256-bit alphanumeric fingerprint for each device. The use of a hash function ensures a negligible collision probability, and the resulting fingerprint's fixed, reduced size enables rapid computational processing.

\subsection{Fingerprint Classifier}

The fingerprint delivered by the Fingerprint Constructor is actionable, for instance, to track a device on the web, only if it is unique. Consequently, the final component of our system is a binary (fingerprint) classifier. Its output, i.e., the dependent variable, is a binary indicator that determines whether a fingerprint is unique (1) or non-unique (0). It receives as input (i.e., independent variables) the \mbR{66 (35)} attributes from the Data Collector for browser (mobile app) samples. 

The Fingerprint Classifier, like any classifier, requires training. To achieve this, we use the \emph{fingerprint} datasets, as introduced in Section~\ref{subsec:ad_campaigns}, comprising \mbR{161k and 52k} real fingerprints from browsers and mobile apps, respectively. Each fingerprint in these datasets is labelled as either unique or non-unique.

The algorithm we employ to build the model for our Fingerprint Classifier is XGBoost (eXtreme Gradient Boosting), which outperformed alternative methods-such as Logistic Regression, Support Vector Machine, Decision Tree, Random Forest, Gradient Boosting Machine, as well as third-party gradient boosting algorithms like LightGBM and CatBoost-in our evaluations. To robustly estimate model performance while maximizing the use of our labelled data (samples with available advertising ID, see Section~\ref{subsec:ad_campaigns}), we adopt a 10-fold stratified cross-validation procedure-an approach to offer an effective balance between bias and variance for moderate-size datasets \cite{arlot2010survey}. 
Specifically, the dataset is randomly partitioned into 10 equal-sized folds, where in each iteration, the model is trained on 9 folds and tested on the remaining fold. This approach ensures that every fingerprint is used once for testing without any overlap between training and validation, thereby preventing potential data leakage.
Moreover, to prevent correlation leakage at the device level, the 10-fold split is also grouped by advertising ID: all samples linked to the same Ad-ID are assigned to the same fold. The Ad-ID is used solely for ground-truth construction and fold assignment and is never included as a model input.

Beyond standard i.i.d. cross-validation, we also assess out-of-distribution (OOD) generalization by (i) leaving one \texttt{\{browser, OS\}} configuration out (LOCO) and (ii) temporal hold-outs across campaigns, always grouping folds and splits by Ad-ID to prevent device-level leakage. Detailed protocols and results are reported in Appendix~\ref{appendix:b}.

Two essential tasks related to data handling are integral to the design of our classifier. Firstly, we apply the k-nearest neighbor (KNN) imputation technique to address structurally missing values—i.e., attributes that are consistently unreported by certain browser or mobile app configurations due to the inherent behaviour or policy of those environments, not accidental failures. Only numerical attributes are imputed, either originally continuous or transformed from high-cardinality categorical variables via frequency encoding. We use $k=5$ and the \texttt{nan\_euclidean} distance metric, which computes distances over the available attributes only, making it well suited for partially missing fingerprints.
Secondly, since XGBoost is primarily designed to work with numerical inputs,\footnote{Note that the latest versions of XGBoost offer experimental support for categorical variables. However, this support is still limited.} we transform categorical variables following best practices: one-hot encoding for low-cardinality attributes and frequency encoding for high-cardinality ones. Frequency-encoded attributes yield continuous numerical values that reflect the statistical distribution of categories in the dataset, making them suitable for distance-based imputation methods.

\begin{table*}[t]
\centering
\caption{\mbR{Details of the ad campaigns used to populate our \emph{fingerprint} datasets.}}
\label{tab:campaigns_table}
\resizebox{\textwidth}{!}{
\begin{tabular}{@{}l|llllllccc@{}}
\toprule
Campaign ID & Time Frame & Year & Ad Source & Type of Device & OS & CPM & \begin{tabular}[c]{@{}l@{}}delivered\\      impressions\end{tabular} & \begin{tabular}[c]{@{}l@{}}collected samples\\      in our server\end{tabular}    & \begin{tabular}[c]{@{}l@{}}samples with\\      advertising ID\end{tabular}    \\ \midrule
adTag\_adF-01                                 & Feb 11-13            & 2022        & webs; apps              & mobile; desktop               & all                       & 0.02€             & 520007               & 381880                   & 157293                   \\ \midrule
adTag\_adF-02                                 & May 22-23            & 2022        & webs; apps              & mobile; desktop               & Apple; Linux              & 0.03€             & 359909               & 302539                   & 86471                    \\ \midrule
adTag\_adF-03                                 & May 27-29            & 2022        & webs; apps              & desktop                       & macOS; Linux              & 0.32€             & 65066                & 53111                    & 28294                    \\ \midrule
adTag\_adF-04                                 & Jun 01-03            & 2022        & webs; apps              & desktop                       & macOS; Linux              & 0.09€             & 791092               & 590832                   & 314750                   \\ \midrule
adTag\_adF-app\_01                            & Jun 06-10            & 2022        & apps                    & mobile                        & Android; iOS              & 0.03€             & 889335               & 594967                   & 292382                   \\ \midrule
adTag\_adF-05                                 & Jun 28-30            & 2022        & webs; apps              & desktop                       & macOS; Linux              & 0.38€             & 25892                & 23766                    & 22241                    \\ \midrule
adTag\_adF-br\_01                             & May 09-13            & 2023        & webs                    & mobile; desktop               & all                       & 0.10€             & 983582               & 733229                   & 327013                   \\ \midrule
adTag\_adF-br\_02                             & Sep 19-24            & 2023        & webs                    & mobile; desktop               & all                       & 0.06€             & 1134013              & 977378                  & 469080                   \\ \midrule
adTag\_adF-app\_02                            & Oct 03-07            & 2023        & apps                    & mobile                        & Android; iOS              & 0.04€             & 634670               & 380042                   & 261414                   \\ \bottomrule
\end{tabular}
}
\end{table*}

Finally, we choose to develop two separate models, one for browsers and another for mobile apps. This decision is driven by the different number of attributes available in each case, \mbR{66 vs. 35}. We also evaluate the option of building different independent models for individual device configurations in our datasets (e.g., a specific model for desktop devices running a Windows OS or for mobile Android apps), but we observe that the performance of \emph{adF} is similar for the general and configuration-specific models, and thus we decide to implement the two referred general models. To further validate the robustness of these models, we compute additional evaluation metrics. The models achieve ROC AUC values of 90\% and 88\%, and balanced accuracy scores of 78\% and 80\% for browser-based and mobile app fingerprints, respectively.

\section{Measurements and Datasets}
\label{subsec:ad_campaigns}

To assess the vulnerability of different device configurations within browser-based environments to web fingerprinting, we deploy the \emph{adF} system, as described earlier, in live ad campaigns. These campaigns are configured within the Sonata Platform, a Demand Side Platform (DSP) operated by TAPTAP Digital. Sonata is a mid-sized DSP that serves tens of millions of daily ads in 15 countries across Europe, Africa, and the Americas. 
The Data Collector retrieves the \mbR{66}  attributes (highlighted in Table \ref{tab:considered_attributes}) from each of the ads delivered through these ad campaigns. Each sample is then fed into the Fingerprint Constructor and the Fingerprint Classifier to obtain the device's fingerprint and determine whether it is unique or not.
In addition, when available, we also collect the device's advertising ID provided by the DSP as a field.

The resulting dataset from these measurements, known as the \emph{raw-browsers} dataset, comprises a total of \mbR{3.06M} data samples collected at different time intervals (February-June 2022, May-June 2023, and September-October 2023). Each data sample includes the following fields:

\noindent 1. device's fingerprint obtained from the Fingerprint Constructor.

\noindent 2. device's advertising ID obtained from the DSP.

\noindent 3. device's type can take two values: mobile or desktop.

\noindent 4. device's OS captures the OS of the device: iOS or Android for mobile devices; Windows, macOS, or Linux for desktop devices.

\noindent\noindent 5. browser represents the browser used by the device (e.g., Chrome, Safari, Firefox). 

\noindent 6. attributes is a data structure including the value of the \mbR{66} attributes.

We collect a second dataset, comprising \mbR{975k} samples from mobile apps, distinct from the browser-based data. This dataset is referred to as the \emph{raw-apps} dataset. The samples contain the same fields as previously described, with specific differences: (1) the field device's type is always mobile in this case; (2) the field device's OS can take only two values, iOS or Android; (3) the field browser identifies, in this case, the WebView of the Android or iOS system; (4) the data structure attributes includes information for \mbR{35} attributes. 

Unfortunately, due to ethical and consistency concerns, we cannot utilize all the samples in the \emph{raw-browsers} and \emph{raw-apps} datasets. On the one hand, we aim to ensure our study does not subject users to risks they are not already exposed to. In our study, the risk primarily involves having a unique ID assigned. Consequently, we exclusively use data samples for which we possess an informed advertising ID, accounting for \mbR{45.88\% and 56.80\% in our \textit{raw-browser} and \textit{raw-apps} datasets, respectively}. On the other hand, among these samples with an informed advertising ID, we only consider those for which the fingerprint value appears at least twice in the dataset. We do this because we cannot assess whether a fingerprint that appears just once is unique or not. The resulting datasets consist of \mbR{161k and 52k} data samples, and we refer to them as the \emph{fingerprint-browsers} and \emph{fingerprint-apps} datasets, respectively. These datasets include two additional fields, which we term \emph{ground-truth uniqueness} and \emph{measured uniqueness}, respectively.

The \emph{ground-truth uniqueness} is a binary variable that signifies whether a fingerprint is genuinely unique in our dataset (1) or not (0). Specifically, if all the samples in any of the \emph{fingerprint} datasets sharing a specific \emph{device's fingerprint} value have the same advertising ID, the \emph{ground-truth uniqueness} variable for these samples is set to 1 (indicating uniqueness), and it is set to 0 (indicating non-uniqueness) otherwise.

We employ the \emph{ground-truth uniqueness} and the \mbR{66} attributes from the samples in the \emph{fingerprint-browsers} dataset to train the XGBoost algorithm and derive the model to be implemented by our Fingerprint Classifier. The model's output for each sample is then assigned to the \emph{measured uniqueness} field of that sample within the \emph{fingerprint-browsers} dataset. For mobile apps, we follow the same procedure, using the \mbR{35} attributes from the \emph{fingerprint-apps} dataset.

\begin{figure}[t]
\centering
\includegraphics[width=\linewidth]{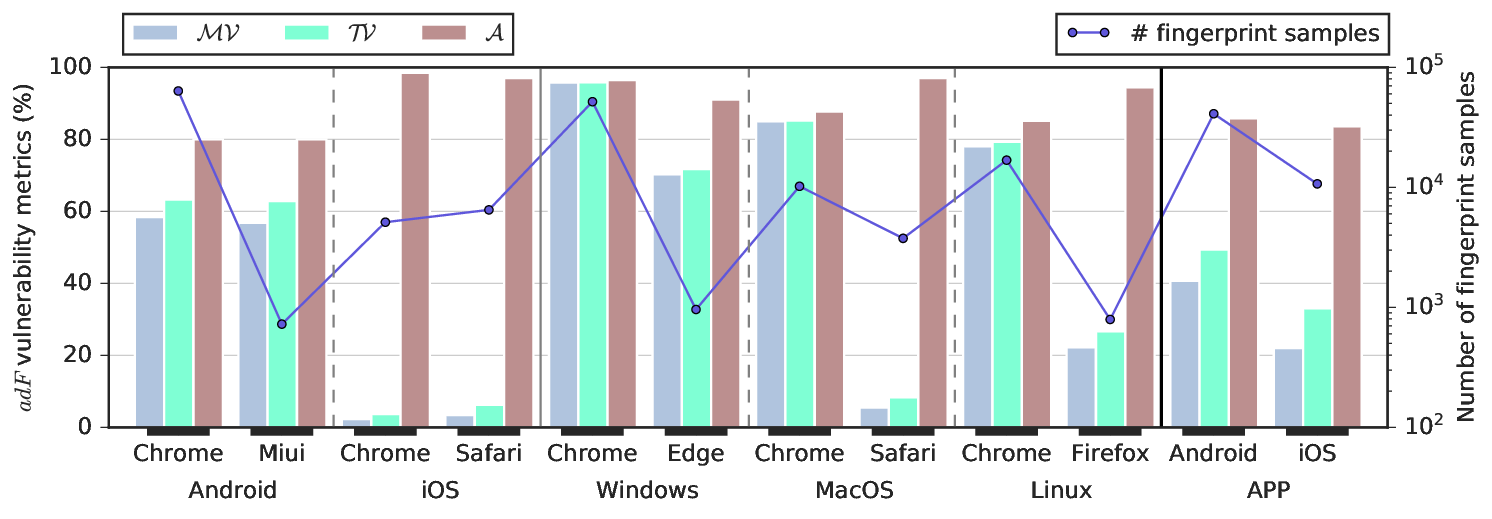}
\caption{\mbR{x-axis: $\mathcal{MV}$, $\mathcal{TV}$ and $\mathcal{A}$ performance of \emph{adF} system for all the considered browser-based device and mobile app configurations; y-axis: number of samples for the browser and mobile app configurations considered for the analyses conducted in the paper.}}
\label{fig:browser_stats_adFP_metrics}
\end{figure}

Finally, it is well-known that the market share of OSes and browsers is imbalanced. On the one hand, \mbR{Chrome holds a two-thirds share of browser installations on both desktop and mobile devices. On the other hand, Windows OS dominates the desktop market with a share over 72.1\%  while the mobile device market is primarily split between Android (71\%) and iOS (28.3\%)}. 
To ensure that our dataset adequately represents the most popular browsers and OSes, we utilize the targeting capabilities provided by Sonata DSP to conduct specific ad campaigns tailored to particular OSes and browsers. Additionally, we run specific campaigns for mobile apps. The details of each campaign can be found in Table \ref{tab:campaigns_table}. In addition, Figure~\ref{fig:browser_stats_adFP_metrics} shows the number of samples in our \emph{fingerprint-browsers} dataset across device configurations accounting for at least \mbR{725} samples. For statistical representativeness reasons, these are the configurations we consider in our analyses in the rest of the paper. In the case of mobile apps, our \emph{fingerprint-apps} dataset includes \mbR{20.66\% (10686) and 79.34\% (41031)} samples from iOS and Android apps, respectively.

\section{Vulnerability \& Accuracy}

In this section, we first define two metrics to assess the vulnerability of a given device configuration to web fingerprinting. In addition, we also introduce the accuracy metric used to assess the performance of \emph{adF} system.

\subsection{Vulnerability Metrics}

\subsubsection{Theoretical Vulnerability (\texorpdfstring{$\mathcal{TV}$}{TV})}

We use \emph{ground-truth uniqueness} from our \emph{fingerprint} datasets to define the theoretical vulnerability ($\mathcal{TV}$) of a given device configuration to web fingerprinting. This is defined as follows:

\begin{equation}
    \mathcal{TV} = \frac{N_{tf}}{N_f}
\end{equation}

Where $N_{f}$ represents the total number of fingerprints and $N_{tf}$ is the number of fingerprints with a \emph{ground-truth uniqueness} equal to 1, for a given device configuration. 
$\mathcal{TV}$ signifies the fraction of fingerprints that are unique in our dataset based on ground-truth data (this is why we use the term \emph{theoretical} to refer to it). 


\subsubsection{Measurable Vulnerability (\texorpdfstring{$\mathcal{MV}$}{MV})}

It is analogous to $\mathcal{TV}$, but with a focus on \emph{measured uniqueness} instead of \emph{ground-truth uniqueness}. In particular, the equation that defines $\mathcal{MV}$ is:

\begin{equation}
    \mathcal{MV} = \frac{N_{mf}}{N_f}
\end{equation}

Where $N_{f}$ represents the total number of fingerprints and $N_{mf}$ is the number of fingerprints with a \emph{measured uniqueness} equal to 1, for a given device configuration.

Hence, while $\mathcal{TV}$ captures the actual vulnerability of a given device configuration to web fingerprinting, $\mathcal{MV}$ reflects the measurable vulnerability using our \emph{adF} system, which is subject to errors (i.e., false positives and false negatives). If our system performs well, then $\mathcal{TV}$ and $\mathcal{MV}$ would exhibit similar values.

\subsection{\emph{adF} system's accuracy}

Our \emph{adF} system essentially  tackles a classification task, aiming to extract device fingerprints and determine their uniqueness. To assess performance, we use a standard classification metric, accuracy ($\mathcal{A}$). Accuracy, in our scenario, represents the proportion of fingerprints correctly classified as unique (true positives) or non-unique (true negatives) by our system, relative to the total number of fingerprints ($N_{f}$). A high accuracy suggests that even without access to advertising ID information, our system can effectively assess the vulnerability of different device configurations to web fingerprinting.

In addition, we can informally assess the performance of our web fingerprinting system by comparing the difference of the $\mathcal{TV}$ and $\mathcal{MV}$ values for different device configurations. 
\section{Results} 
\label{sec:results}

In this section, we first analyze the performance of \emph{adF} system. 
Afterward, we analyze the vulnerability of prevalent device configurations formed by the combination of device type, operating system, and browser. Finally, we also analyze the vulnerability of mobile apps to web fingerprinting.

\subsection{Performance of our \emph{adF} system}

Figure \ref{fig:browser_stats_adFP_metrics} displays the values of $\mathcal{TV}$, $\mathcal{MV}$, and $\mathcal{A}$ for each configured mobile app and browser-based setup. \emph{adF} consistently achieves high accuracy, with $\mathcal{A}$ values exceeding \mbR{80\%} for all configurations. Moreover, $\mathcal{MV}$ and $\mathcal{TV}$ exhibit similar values across the configurations. Indeed, the maximum difference is only 11.02 percentage points for iOS apps. 
As such, we can confidently assert that \emph{adF} is a reliable tool for assessing the vulnerability of different device configurations to web fingerprinting, even in the absence of ground-truth data like advertising IDs.

\subsection{Vulnerability of browser-based device configurations to web fingerprinting}

The values of $\mathcal{TV}$ and $\mathcal{MV}$ presented in Figure \ref{fig:browser_stats_adFP_metrics} serve to assess the vulnerability of each of the browser-based configurations to web fingerprinting. Given the similarity between $\mathcal{TV}$ and $\mathcal{MV}$, we discuss results based solely on $\mathcal{MV}$ for the remainder of this section.

The first conclusion we extract from the results in Figure \ref{fig:browser_stats_adFP_metrics} is that mobile devices exhibit lower vulnerability to web fingerprinting when compared to desktop devices for a given browser (Chrome or Safari). For instance, in desktop devices running macOS, Safari's $\mathcal{MV}$ is \mbR{5.5\%}, while in mobile devices running iOS, it is \mbR{3.4\%}. Likewise, Chrome offers lower $\mathcal{MV}$ in mobile devices compared to desktops, regardless of the OS. 

Second, the choice of operating system (OS) significantly influences defense against web fingerprinting. 
On mobile devices, regardless of the browser, iOS users enjoy better protection against web fingerprinting than their Android counterparts. For desktop devices, macOS and Linux outperform Windows in shielding against web fingerprinting. When considering the common browser, Chrome, across these OS categories, $\mathcal{MV}$ rates are \mbR{$\lbrace$95.7\%, 85\%, 78.1\%$\rbrace$ for $\lbrace$Windows, macOS, Linux$\rbrace$}. 
However, unlike mobile devices, a single OS does not consistently offer superior protection regardless of the browser. For example, Chrome users on macOS and Edge users on Windows are more vulnerable compared to Firefox users on Linux. 

Finally, when examining browsers individually, we find that Chrome ranks as the least secure option (except in the case of iOS). In contrast, Safari and Firefox provide notably superior protection against web fingerprinting. Note that we gather a few hundred samples from Firefox on both Windows and macOS, suggesting Firefox's strong performance also on these operating systems. \mbR{To better understand the outstanding performance of Chrome on iOS, we refer the reader to our analysis in Section \ref{sec:solution}. This result is due to the low cardinality and entropy of attributes accessible in Chrome for iOS devices, which does not occur in other OSes.}

\vspace{2mm}
\noindent{Putting the results into context:}
\vspace{1mm}

The configurations in Figure \ref{fig:browser_stats_adFP_metrics} correspond to the most prevalent setups in the current web landscape, accounting for over \mbR{90\%} of all devices.

Our findings reveal significant variability in the vulnerability of these configurations to web fingerprinting. 
To put these results into perspective, it is essential to consider the market share, which is notably skewed, both in terms of OSes and browsers\footnote{Market share statistics are based on data from https://gs.statcounter.com/, covering Feb 2022 to Oct 2023, providing a contemporary context to our dataset's time-frame.}.

\mbR{In the mobile ecosystem, 71.01\% and 28.30\% of the devices run on Android and iOS, respectively. Regarding browsers, 64.37\% of devices employ Chrome, and 24.98\% opt for Safari. By combining OSes and browsers' market share, we identify three configurations that cover 95\% of the mobile devices' market share: \texttt{\{Chrome, Android\}} $\sim$67.7\%, \texttt{\{Safari, iOS\}} $\sim$25\%, and \texttt{\{Chrome, iOS\}} $\sim$3.3\%. 
When we extrapolate the $\mathcal{MV}$ values presented in Figure \ref{fig:browser_stats_adFP_metrics} to this market share, we observe that web fingerprinting is feasible for at least \mbR{40\%} of mobile devices when exposed to ads.}

\mbR{In the context of the desktop ecosystem, in terms of OSes, Windows, macOS, and Linux cover 72.09\%, 16.85\%, and 2.77\% of the market share, respectively. When it comes to browsers, Chrome dominates with \mbR{65.59\%}, Edge follows with \mbR{10.66\%}, Safari has \mbR{10.60\%}, and Firefox stands at \mbR{6.95\%}. 
Therefore, the most prevalent desktop configurations are: \texttt{\{Chrome, Windows\}} $\sim$54.5\%, \texttt{\{Edge, Windows\}} $\sim$10.7\%, \texttt{\{Safari, macOS\}} $\sim$10.6\%, \texttt{\{Chrome, macOS\}} $\sim$5.7\%, \texttt{\{Chrome, Linux\}} $\sim$1.2\%, and \texttt{\{Firefox, Linux\}} $\sim$1.2\%.
When extrapolating the results from Figure \ref{fig:browser_stats_adFP_metrics} to this specific market share, we find that the prevalence of the \texttt{\{Chrome, Windows\}} configuration leads to a situation where at least 66\% of desktop devices can be effectively fingerprinted from ads.}

\subsection{Mobile apps}

Most mobile apps operate on top of Android or iOS WebView; therefore, they can be exploited to implement web fingerprinting. However, to the best of the authors' knowledge, there is no previous empirical study that confirms if this practice is viable. Figure \ref{fig:browser_stats_adFP_metrics} shows the value of $\mathcal{TV}$, $\mathcal{MV}$ and $\mathcal{A}$ for mobile apps on \emph{Android} vs. \emph{iOS}.

First, we confirm that web fingerprinting is a viable practice in mobile applications. Furthermore, our experiments also confirm that \emph{adF} is a valid tool to assess the vulnerability of mobile applications to web fingerprinting. In particular, $\mathcal{A}$ is higher than \mbR{83\%} for both iOS and Android apps.

Second, Android apps are more exposed to web fingerprinting compared to iOS. In particular, when considering $\mathcal{MV}$, we find that \mbR{40.7\%} of Android users and \mbR{22\%} of iOS users can be uniquely fingerprinted using mobile apps' ads. If we extrapolate these results to the actual market share of Android and iOS \mbR{(71.01\% and 28.30\%, respectively), we conclude that 35\%} of mobile devices can be uniquely identified using web fingerprinting techniques in mobile apps.

Finally, when comparing app versus browser vulnerability on mobile web, we find that while on Android, device vulnerability is significantly lower in mobile apps than browsers, on iOS, apps exhibit a notably higher vulnerability to web fingerprinting compared to browsers. This is a relevant finding for iOS mobile app developers who should consider when striving to enhance the security offered by their apps.


\section{Methodology Limitations}

Our dataset is limited in device coverage, so our claim of fingerprint uniqueness is confined to our dataset's scope, and we cannot claim the global uniqueness of such fingerprints. This limitation stems primarily from our restricted budget for ad campaigns, amounting to \mbR{410.27 €}. Also, ethical considerations led us to exclude almost half of the samples without an advertising ID in our dataset. Nevertheless, despite these constraints, our dataset's size (see Section \ref{sec:results}) is comparable to previous works in the literature \cite{gomezBoix2018hiding, andriamilanto2021largescale}. It confirms that \emph{adF} provides high accuracy, establishing it as a reliable system for evaluating device configuration vulnerability to web fingerprinting. Furthermore, our results highlight the significant variability in protection among different device configurations.

We also acknowledge that our methodology faces different issues that impede collecting the required data from every single ad executing our Data Collector's \emph{adTag}. Some of these issues include: browsers running ad-blockers or disabling JavaScript, outdated browsers, limited browser API support, network problems in the connection between the ad and our server, etc. Moreover, the runtime required by our \emph{adTag} 
represents another source of losses. Our \emph{adTag} runs as long as the web-page (mobile app) where the ad is rendered is open on the user's device. Hence, once the user closes the web-page (mobile app), the \emph{adTag} execution is stopped. Callejo et al. \cite{callejo2017opportunities} estimated that \mbR{75\%} of ads displayed on PCs and mobile devices remain active for at least \mbR{15 and 8} seconds, respectively. Although our \emph{adTag} has an overall estimated runtime in the order of hundreds of milliseconds (depending on the browser and OS), we must also consider the rendering time of the website itself. All the mentioned causes lead to an average loss of \mbR{~19\%} ad impressions from web-pages and of \mbR{~33\%} ad impressions from mobile apps by comparing the campaign's number of impressions reported by the DSP and the total samples collected in our back-end server (See Table~\ref{tab:campaigns_table}).

\section{Causes of Web Fingerprinting Vulnerability}
\label{sec:atributes_analysis}

Different device configurations offer a significantly different vulnerability to web fingerprinting. In this section, we dissect the main factors that define the probability of a fingerprint being unique and, by extension, the resilience of a given device configuration to web fingerprinting.

\subsection{Reported attribute count (\texorpdfstring{$|N|$}{N})}

An obvious factor that influences the resilience of a given device configuration (expressed through $\mathcal{TV}$ or $\mathcal{MV}$) is the length of the vector forming the fingerprint ($|N|$) by such configuration to third-parties. For instance, in desktop devices, \texttt{\{Safari, macOS\}} provides information for only \mbR{39 out of the 66} attributes considered by \emph{adF} 
whereas Firefox reports (regardless of the OS) data for \mbR{61 of these 66} attributes. Finally, Chrome provides, in any OS, information for all \mbR{66} attributes. Hence, in principle, Chrome users are more exposed to web fingerprinting since it is simpler to generate unique fingerprints using a larger pool of attributes.

Longer vectors generally result in a higher probability of acquiring unique fingerprints. In other words, device configurations that provide a greater number of attributes are typically more vulnerable to web fingerprinting.
However, this is not the sole factor influencing the resilience to web fingerprinting. For instance, in the case of desktop configurations, if we use Safari's anti-fingerprint settings in Chrome samples on Windows (i.e., we do not consider in Chrome those attributes not reported by Safari), the $\mathcal{MV}$ ($\mathcal{TV}$) for \texttt{\{Chrome, Windows\}} devices become \mbR{53.71\% (60.35\%)}. This represents an important improvement compared to the original vulnerability measured for \texttt{\{Chrome, Windows\}} configuration, \mbR{95.74\% (95.81\%)}. However, it is still far from \texttt{\{Safari, macOS\}} results \mbR{($\mathcal{MV}$ = 5.48\%, $\mathcal{TV}$ = 8.36\%)}. This example shows that two device configurations reporting exactly the same attributes present significantly different web fingerprinting vulnerabilities. Hence, there must be other factors influencing the web fingerprinting vulnerability.



\begin{table*}[t!]
\centering
\caption{\mbR{Discrimination power of attributes. For each attribute and device configuration, we report: (i) whether the attribute is reported or not, (ii) the cardinality ($|S|$), (iii) the normalized entropy ({$h$}). We highlight in bold those attributes whose reporting we propose to block in our solution ($|S|>$ 25 and a $h\geq$ 0.10)}.}
\label{tab:featuresAttributes}
\resizebox{\textwidth}{!}{%
\begin{tabular}{|l|cccccccc|cccccccccccc|cccc|}
\hline
\multicolumn{1}{|c|}{\multirow{4}{*}{Attributes}}  & \multicolumn{8}{c|}{MOBILE}                                                                                                                & \multicolumn{12}{c|}{DESKTOP}                                                                                                                                                                                         & \multicolumn{4}{c|}{APP}                                                                       \\ \cline{2-25} 
\multicolumn{1}{|c|}{}                             & \multicolumn{4}{c|}{Android}                                            & \multicolumn{4}{c|}{iOS}                                         & \multicolumn{4}{c|}{Windows}                                            & \multicolumn{4}{c|}{macOS}                                              & \multicolumn{4}{c|}{GNU/Linux}                                    & \multicolumn{2}{c|}{\multirow{2}{*}{Android}}      & \multicolumn{2}{c|}{\multirow{2}{*}{iOS}} \\ \cline{2-21}
\multicolumn{1}{|c|}{}                             & \multicolumn{2}{c|}{Chrome}        & \multicolumn{2}{c|}{MIUI}          & \multicolumn{2}{c|}{Chrome}        & \multicolumn{2}{c|}{Safari} & \multicolumn{2}{c|}{Chrome}        & \multicolumn{2}{c|}{Edge}          & \multicolumn{2}{c|}{Chrome}        & \multicolumn{2}{c|}{Safari}        & \multicolumn{2}{c|}{Chrome}        & \multicolumn{2}{c|}{Firefox} & \multicolumn{2}{c|}{} & \multicolumn{2}{c|}{}       \\ \cline{2-25} 
\multicolumn{1}{|c|}{}                             & $|S|$ & \multicolumn{1}{c|}{$h$} & $|S|$ & \multicolumn{1}{c|}{$h$} & $|S|$ & \multicolumn{1}{c|}{$h$} & $|S|$        & $h$        & $|S|$ & \multicolumn{1}{c|}{$h$} & $|S|$ & \multicolumn{1}{c|}{$h$} & $|S|$ & \multicolumn{1}{c|}{$h$} & $|S|$ & \multicolumn{1}{c|}{$h$} & $|S|$ & \multicolumn{1}{c|}{$h$} & $|S|$         & $h$        & $|S|$                 & \multicolumn{1}{c|}{$h$} & $|S|$                       & $h$       \\ \hline
UserAgent                                          & 6946  & \multicolumn{1}{c|}{0.39}  & 165   & \multicolumn{1}{c|}{0.44}  & 602   & \multicolumn{1}{c|}{0.57}  & 131          & 0.38         & 374   & \multicolumn{1}{c|}{0.28}  & 56    & \multicolumn{1}{c|}{0.34}  & 141   & \multicolumn{1}{c|}{0.25}  & 72    & \multicolumn{1}{c|}{0.36}  & 216   & \multicolumn{1}{c|}{0.21}  & 82            & 0.46         & 5050                  & \multicolumn{1}{c|}{0.73}  & 106                         & 0.33        \\ \hline
\textbf{CPU cores}                & 8     & \multicolumn{1}{c|}{0.02}  & 2     & \multicolumn{1}{c|}{0.01}  & 4     & \multicolumn{1}{c|}{0.03}  & 4            & 0.05         & 24    & \multicolumn{1}{c|}{0.14}  & 9     & \multicolumn{1}{c|}{0.22}  & 13    & \multicolumn{1}{c|}{0.14}  & 4     & \multicolumn{1}{c|}{0.12}  & 24    & \multicolumn{1}{c|}{0.13}  & 9             & 0.23         & 6                     & \multicolumn{1}{c|}{0.02}  & 2                           & 0.03        \\ \hline
\textbf{Device memory}            & 6     & \multicolumn{1}{c|}{0.09}  & 4     & \multicolumn{1}{c|}{0.10}  & 1     & \multicolumn{1}{c|}{0.00}  & 1            & 0.00         & 6     & \multicolumn{1}{c|}{0.08}  & 5     & \multicolumn{1}{c|}{0.09}  & 4     & \multicolumn{1}{c|}{0.03}  & 1     & \multicolumn{1}{c|}{0.00}  & 6     & \multicolumn{1}{c|}{0.11}  & 1             & 0.00         & 5                     & \multicolumn{1}{c|}{0.10}  & 1                           & 0.00        \\ \hline
Screen: color depth                                & 1     & \multicolumn{1}{c|}{0.00}  & 1     & \multicolumn{1}{c|}{0.00}  & 2     & \multicolumn{1}{c|}{0.02}  & 3            & 0.02         & 4     & \multicolumn{1}{c|}{0.00}  & 3     & \multicolumn{1}{c|}{0.01}  & 3     & \multicolumn{1}{c|}{0.08}  & 2     & \multicolumn{1}{c|}{0.08}  & 4     & \multicolumn{1}{c|}{0.00}  & 2             & 0.02         & --                    & \multicolumn{1}{c|}{--}    & --                          & --          \\ \hline
Screen: pixel left                                 & 4     & \multicolumn{1}{c|}{0.00}  & 1     & \multicolumn{1}{c|}{0.00}  & 1     & \multicolumn{1}{c|}{0.00}  & 2            & 0.00         & 1615  & \multicolumn{1}{c|}{0.16}  & 74    & \multicolumn{1}{c|}{0.17}  & 735   & \multicolumn{1}{c|}{0.21}  & 286   & \multicolumn{1}{c|}{0.15}  & 609   & \multicolumn{1}{c|}{0.17}  & 101           & 0.32         & --                    & \multicolumn{1}{c|}{--}    & --                          & --          \\ \hline
Screen: orientation angle                          & 4     & \multicolumn{1}{c|}{0.05}  & 3     & \multicolumn{1}{c|}{0.05}  & 4     & \multicolumn{1}{c|}{0.08}  & 4            & 0.10         & 4     & \multicolumn{1}{c|}{0.00}  & 4     & \multicolumn{1}{c|}{0.01}  & 2     & \multicolumn{1}{c|}{0.00}  & 5     & \multicolumn{1}{c|}{0.14}  & 4     & \multicolumn{1}{c|}{0.05}  & 1             & 0.00         & 4                     & \multicolumn{1}{c|}{0.00}  & 5                           & 0.06        \\ \hline
Screen: orientation type                           & 4     & \multicolumn{1}{c|}{0.05}  & 3     & \multicolumn{1}{c|}{0.05}  & 4     & \multicolumn{1}{c|}{0.08}  & 4            & 0.10         & 4     & \multicolumn{1}{c|}{0.00}  & 4     & \multicolumn{1}{c|}{0.00}  & 3     & \multicolumn{1}{c|}{0.00}  & 5     & \multicolumn{1}{c|}{0.14}  & 4     & \multicolumn{1}{c|}{0.07}  & 1             & 0.00         & --                    & \multicolumn{1}{c|}{--}    & --                          & --          \\ \hline
Battery status: charging                           & 3     & \multicolumn{1}{c|}{0.04}  & 3     & \multicolumn{1}{c|}{0.05}  & 1     & \multicolumn{1}{c|}{0.00}  & 1            & 0.00         & 3     & \multicolumn{1}{c|}{0.03}  & 3     & \multicolumn{1}{c|}{0.05}  & 3     & \multicolumn{1}{c|}{0.08}  & 1     & \multicolumn{1}{c|}{0.00}  & 3     & \multicolumn{1}{c|}{0.07}  & 1             & 0.00         & --                    & \multicolumn{1}{c|}{--}    & --                          & --          \\ \hline
Simultaneous touch points                          & 4     & \multicolumn{1}{c|}{0.02}  & 1     & \multicolumn{1}{c|}{0.00}  & 3     & \multicolumn{1}{c|}{0.00}  & 4            & 0.00         & 19    & \multicolumn{1}{c|}{0.02}  & 6     & \multicolumn{1}{c|}{0.05}  & 2     & \multicolumn{1}{c|}{0.00}  & 3     & \multicolumn{1}{c|}{0.08}  & 8     & \multicolumn{1}{c|}{0.09}  & 2             & 0.00         & 2                     & \multicolumn{1}{c|}{0.03}  & 2                           & 0.01        \\ \hline
\textbf{Media devices}            & 39    & \multicolumn{1}{c|}{0.17}  & 6     & \multicolumn{1}{c|}{0.01}  & 5     & \multicolumn{1}{c|}{0.08}  & 17           & 0.10         & 66    & \multicolumn{1}{c|}{0.14}  & 8     & \multicolumn{1}{c|}{0.16}  & 8     & \multicolumn{1}{c|}{0.05}  & 12    & \multicolumn{1}{c|}{0.10}  & 43    & \multicolumn{1}{c|}{0.21}  & 28            & 0.21         & 22                    & \multicolumn{1}{c|}{0.18}  & 5                           & 0.08        \\ \hline
\textbf{Languages}                & 2748  & \multicolumn{1}{c|}{0.31}  & 22    & \multicolumn{1}{c|}{0.13}  & 27    & \multicolumn{1}{c|}{0.15}  & 39           & 0.18         & 1638  & \multicolumn{1}{c|}{0.20}  & 55    & \multicolumn{1}{c|}{0.24}  & 822   & \multicolumn{1}{c|}{0.36}  & 43    & \multicolumn{1}{c|}{0.18}  & 1355  & \multicolumn{1}{c|}{0.39}  & 40            & 0.25         & 285                   & \multicolumn{1}{c|}{0.17}  & 42                          & 0.20        \\ \hline
PDF viewer enabled                                 & 3     & \multicolumn{1}{c|}{0.01}  & 2     & \multicolumn{1}{c|}{0.05}  & 3     & \multicolumn{1}{c|}{0.08}  & 3            & 0.08         & 3     & \multicolumn{1}{c|}{0.04}  & 3     & \multicolumn{1}{c|}{0.01}  & 3     & \multicolumn{1}{c|}{0.01}  & 3     & \multicolumn{1}{c|}{0.09}  & 3     & \multicolumn{1}{c|}{0.08}  & 3             & 0.09         & --                    & \multicolumn{1}{c|}{--}    & --                          & --          \\ \hline
\textbf{User Permissions state}   & 196   & \multicolumn{1}{c|}{0.13}  & 1     & \multicolumn{1}{c|}{0.00}  & 6     & \multicolumn{1}{c|}{0.18}  & 22           & 0.20         & 113   & \multicolumn{1}{c|}{0.13}  & 21    & \multicolumn{1}{c|}{0.17}  & 58    & \multicolumn{1}{c|}{0.15}  & 19    & \multicolumn{1}{c|}{0.19}  & 155   & \multicolumn{1}{c|}{0.22}  & 16            & 0.25         & --                    & \multicolumn{1}{c|}{--}    & --                          & --          \\ \hline
available height                                   & 456   & \multicolumn{1}{c|}{0.32}  & 41    & \multicolumn{1}{c|}{0.31}  & 30    & \multicolumn{1}{c|}{0.29}  & 32           & 0.25         & 549   & \multicolumn{1}{c|}{0.26}  & 79    & \multicolumn{1}{c|}{0.42}  & 582   & \multicolumn{1}{c|}{0.52}  & 310   & \multicolumn{1}{c|}{0.45}  & 590   & \multicolumn{1}{c|}{0.44}  & 206           & 0.63         & 227                   & \multicolumn{1}{c|}{0.33}  & 20                          & 0.24        \\ \hline
available left                                     & 1     & \multicolumn{1}{c|}{0.00}  & 1     & \multicolumn{1}{c|}{0.00}  & 1     & \multicolumn{1}{c|}{0.00}  & 3            & 0.00         & 221   & \multicolumn{1}{c|}{0.04}  & 17    & \multicolumn{1}{c|}{0.04}  & 199   & \multicolumn{1}{c|}{0.07}  & 30    & \multicolumn{1}{c|}{0.01}  & 166   & \multicolumn{1}{c|}{0.14}  & 60            & 0.27         & --                    & \multicolumn{1}{c|}{--}    & --                          & --          \\ \hline
available top                                      & 1     & \multicolumn{1}{c|}{0.00}  & 1     & \multicolumn{1}{c|}{0.00}  & 1     & \multicolumn{1}{c|}{0.00}  & 3            & 0.00         & 484   & \multicolumn{1}{c|}{0.03}  & 13    & \multicolumn{1}{c|}{0.02}  & 181   & \multicolumn{1}{c|}{0.18}  & 13    & \multicolumn{1}{c|}{0.11}  & 170   & \multicolumn{1}{c|}{0.12}  & 38            & 0.28         & --                    & \multicolumn{1}{c|}{--}    & --                          & --          \\ \hline
available width                                    & 445   & \multicolumn{1}{c|}{0.24}  & 30    & \multicolumn{1}{c|}{0.17}  & 30    & \multicolumn{1}{c|}{0.26}  & 33           & 0.21         & 502   & \multicolumn{1}{c|}{0.20}  & 57    & \multicolumn{1}{c|}{0.31}  & 171   & \multicolumn{1}{c|}{0.24}  & 79    & \multicolumn{1}{c|}{0.35}  & 541   & \multicolumn{1}{c|}{0.37}  & 156           & 0.51         & 124                   & \multicolumn{1}{c|}{0.20}  & 16                          & 0.19        \\ \hline
full screen enabled                                & 2     & \multicolumn{1}{c|}{0.06}  & 2     & \multicolumn{1}{c|}{0.10}  & 3     & \multicolumn{1}{c|}{0.01}  & 1            & 0.00         & 3     & \multicolumn{1}{c|}{0.05}  & 2     & \multicolumn{1}{c|}{0.10}  & 2     & \multicolumn{1}{c|}{0.07}  & 3     & \multicolumn{1}{c|}{0.11}  & 3     & \multicolumn{1}{c|}{0.07}  & 3             & 0.10         & 2                     & \multicolumn{1}{c|}{0.06}  & 1                           & 0.00        \\ \hline
\textbf{Storage: quota}           & 3650  & \multicolumn{1}{c|}{0.60}  & 198   & \multicolumn{1}{c|}{0.67}  & 12    & \multicolumn{1}{c|}{0.03}  & 18           & 0.03         & 23271 & \multicolumn{1}{c|}{0.81}  & 743   & \multicolumn{1}{c|}{0.93}  & 772   & \multicolumn{1}{c|}{0.39}  & 20    & \multicolumn{1}{c|}{0.02}  & 4664  & \multicolumn{1}{c|}{0.74}  & 91            & 0.21         & 1787                  & \multicolumn{1}{c|}{0.62}  & 3                           & 0.06        \\ \hline
\textbf{navigator properties}     & 72    & \multicolumn{1}{c|}{0.18}  & 5     & \multicolumn{1}{c|}{0.12}  & 25    & \multicolumn{1}{c|}{0.24}  & 37           & 0.20         & 97    & \multicolumn{1}{c|}{0.16}  & 18    & \multicolumn{1}{c|}{0.21}  & 44    & \multicolumn{1}{c|}{0.17}  & 29    & \multicolumn{1}{c|}{0.27}  & 120   & \multicolumn{1}{c|}{0.22}  & 24            & 0.25         & 31                    & \multicolumn{1}{c|}{0.10}  & 29                          & 0.25        \\ \hline
Plugins                                            & 1     & \multicolumn{1}{c|}{0.00}  & 1     & \multicolumn{1}{c|}{0.00}  & 2     & \multicolumn{1}{c|}{0.08}  & 2            & 0.08         & 51    & \multicolumn{1}{c|}{0.05}  & 4     & \multicolumn{1}{c|}{0.01}  & 8     & \multicolumn{1}{c|}{0.02}  & 21    & \multicolumn{1}{c|}{0.17}  & 47    & \multicolumn{1}{c|}{0.09}  & 4             & 0.09         & --                    & \multicolumn{1}{c|}{--}    & --                          & --          \\ \hline
Cookie enabled                                     & 2     & \multicolumn{1}{c|}{0.00}  & 2     & \multicolumn{1}{c|}{0.01}  & 2     & \multicolumn{1}{c|}{0.08}  & 2            & 0.07         & 2     & \multicolumn{1}{c|}{0.00}  & 2     & \multicolumn{1}{c|}{0.01}  & 2     & \multicolumn{1}{c|}{0.00}  & 2     & \multicolumn{1}{c|}{0.07}  & 2     & \multicolumn{1}{c|}{0.00}  & 2             & 0.07         & --                    & \multicolumn{1}{c|}{--}    & --                          & --          \\ \hline
MIME type                                          & 1     & \multicolumn{1}{c|}{0.00}  & 1     & \multicolumn{1}{c|}{0.00}  & 2     & \multicolumn{1}{c|}{0.08}  & 2            & 0.08         & 21    & \multicolumn{1}{c|}{0.05}  & 4     & \multicolumn{1}{c|}{0.01}  & 7     & \multicolumn{1}{c|}{0.02}  & 7     & \multicolumn{1}{c|}{0.12}  & 15    & \multicolumn{1}{c|}{0.09}  & 6             & 0.10         & --                    & \multicolumn{1}{c|}{--}    & --                          & --          \\ \hline
Time zone offset                                   & 20    & \multicolumn{1}{c|}{0.15}  & 7     & \multicolumn{1}{c|}{0.12}  & 13    & \multicolumn{1}{c|}{0.13}  & 19           & 0.15         & 24    & \multicolumn{1}{c|}{0.08}  & 9     & \multicolumn{1}{c|}{0.17}  & 15    & \multicolumn{1}{c|}{0.15}  & 14    & \multicolumn{1}{c|}{0.11}  & 14    & \multicolumn{1}{c|}{0.09}  & 9             & 0.14         & 15                    & \multicolumn{1}{c|}{0.13}  & 12                          & 0.18        \\ \hline
\textbf{Canvas}                   & 1360  & \multicolumn{1}{c|}{0.46}  & 70    & \multicolumn{1}{c|}{0.45}  & 90    & \multicolumn{1}{c|}{0.31}  & 311          & 0.35         & 619   & \multicolumn{1}{c|}{0.33}  & 134   & \multicolumn{1}{c|}{0.57}  & 247   & \multicolumn{1}{c|}{0.39}  & 236   & \multicolumn{1}{c|}{0.47}  & 1078  & \multicolumn{1}{c|}{0.50}  & 182           & 0.55         & 380                   & \multicolumn{1}{c|}{0.37}  & 102                         & 0.31        \\ \hline
\textbf{Fonts}                    & 14    & \multicolumn{1}{c|}{0.05}  & 2     & \multicolumn{1}{c|}{0.06}  & 13    & \multicolumn{1}{c|}{0.10}  & 17           & 0.11         & 8359  & \multicolumn{1}{c|}{0.56}  & 361   & \multicolumn{1}{c|}{0.71}  & 1934  & \multicolumn{1}{c|}{0.52}  & 78    & \multicolumn{1}{c|}{0.24}  & 1382  & \multicolumn{1}{c|}{0.36}  & 299           & 0.70         & 2                     & \multicolumn{1}{c|}{0.00}  & 7                           & 0.11        \\ \hline
\textbf{Bluetooth   availability} & 3     & \multicolumn{1}{c|}{0.06}  & 2     & \multicolumn{1}{c|}{0.05}  & 1     & \multicolumn{1}{c|}{0.00}  & 1            & 0.00         & 3     & \multicolumn{1}{c|}{0.08}  & 3     & \multicolumn{1}{c|}{0.10}  & 3     & \multicolumn{1}{c|}{0.08}  & 1     & \multicolumn{1}{c|}{0.00}  & 3     & \multicolumn{1}{c|}{0.10}  & 1             & 0.00         & --                    & \multicolumn{1}{c|}{--}    & --                          & --          \\ \hline
\textbf{WebGL   Extensions}       & 86    & \multicolumn{1}{c|}{0.23}  & 15    & \multicolumn{1}{c|}{0.20}  & 10    & \multicolumn{1}{c|}{0.15}  & 17           & 0.17         & 52    & \multicolumn{1}{c|}{0.10}  & 13    & \multicolumn{1}{c|}{0.16}  & 16    & \multicolumn{1}{c|}{0.05}  & 23    & \multicolumn{1}{c|}{0.29}  & 126   & \multicolumn{1}{c|}{0.31}  & 30            & 0.23         & 57                    & \multicolumn{1}{c|}{0.22}  & 8                           & 0.12        \\ \hline
Audio formats: AACP                                & 1     & \multicolumn{1}{c|}{0.00}  & 1     & \multicolumn{1}{c|}{0.00}  & 2     & \multicolumn{1}{c|}{0.00}  & 2            & 0.00         & 2     & \multicolumn{1}{c|}{0.00}  & 1     & \multicolumn{1}{c|}{0.00}  & 1     & \multicolumn{1}{c|}{0.00}  & 2     & \multicolumn{1}{c|}{0.00}  & 1     & \multicolumn{1}{c|}{0.00}  & 3             & 0.02         & --                    & \multicolumn{1}{c|}{--}    & --                          & --          \\ \hline
Audio formats: ACC                                 & 2     & \multicolumn{1}{c|}{0.00}  & 1     & \multicolumn{1}{c|}{0.00}  & 3     & \multicolumn{1}{c|}{0.00}  & 3            & 0.00         & 3     & \multicolumn{1}{c|}{0.00}  & 1     & \multicolumn{1}{c|}{0.00}  & 2     & \multicolumn{1}{c|}{0.00}  & 3     & \multicolumn{1}{c|}{0.01}  & 2     & \multicolumn{1}{c|}{0.00}  & 3             & 0.01         & 1                     & \multicolumn{1}{c|}{0.00}  & 2                           & 0.01        \\ \hline
Audio cxt: base latency                            & 116   & \multicolumn{1}{c|}{0.22}  & 9     & \multicolumn{1}{c|}{0.22}  & 8     & \multicolumn{1}{c|}{0.06}  & 11           & 0.05         & 18    & \multicolumn{1}{c|}{0.02}  & 3     & \multicolumn{1}{c|}{0.03}  & 9     & \multicolumn{1}{c|}{0.09}  & 13    & \multicolumn{1}{c|}{0.11}  & 76    & \multicolumn{1}{c|}{0.20}  & 5             & 0.04         & 70                    & \multicolumn{1}{c|}{0.25}  & 7                           & 0.05        \\ \hline
Audio cxt: max channel count                       & 1     & \multicolumn{1}{c|}{0.00}  & 1     & \multicolumn{1}{c|}{0.00}  & 4     & \multicolumn{1}{c|}{0.02}  & 4            & 0.03         & 5     & \multicolumn{1}{c|}{0.02}  & 5     & \multicolumn{1}{c|}{0.02}  & 14    & \multicolumn{1}{c|}{0.01}  & 7     & \multicolumn{1}{c|}{0.03}  & 5     & \multicolumn{1}{c|}{0.00}  & 5             & 0.02         & --                    & \multicolumn{1}{c|}{--}    & --                          & --          \\ \hline
Audio cxt: sample rate                             & 4     & \multicolumn{1}{c|}{0.01}  & 2     & \multicolumn{1}{c|}{0.00}  & 6     & \multicolumn{1}{c|}{0.05}  & 7            & 0.04         & 10    & \multicolumn{1}{c|}{0.03}  & 5     & \multicolumn{1}{c|}{0.04}  & 9     & \multicolumn{1}{c|}{0.08}  & 10    & \multicolumn{1}{c|}{0.09}  & 7     & \multicolumn{1}{c|}{0.07}  & 3             & 0.09         & 2                     & \multicolumn{1}{c|}{0.01}  & 6                           & 0.04        \\ \hline
Audio cxt: state                                   & 2     & \multicolumn{1}{c|}{0.05}  & 2     & \multicolumn{1}{c|}{0.03}  & 2     & \multicolumn{1}{c|}{0.00}  & 2            & 0.00         & 2     & \multicolumn{1}{c|}{0.06}  & 2     & \multicolumn{1}{c|}{0.09}  & 2     & \multicolumn{1}{c|}{0.07}  & 2     & \multicolumn{1}{c|}{0.00}  & 2     & \multicolumn{1}{c|}{0.07}  & 2             & 0.00         & --                    & \multicolumn{1}{c|}{--}    & --                          & --          \\ \hline
\textbf{WebGL (Rend - Param)}     & 525   & \multicolumn{1}{c|}{0.46}  & 76    & \multicolumn{1}{c|}{0.52}  & 10    & \multicolumn{1}{c|}{0.09}  & 18           & 0.11         & 2038  & \multicolumn{1}{c|}{0.47}  & 310   & \multicolumn{1}{c|}{0.72}  & 319   & \multicolumn{1}{c|}{0.46}  & 52    & \multicolumn{1}{c|}{0.24}  & 2360  & \multicolumn{1}{c|}{0.65}  & 137           & 0.54         & 234                   & \multicolumn{1}{c|}{0.40}  & 7                           & 0.05        \\ \hline
\end{tabular}%
}
\end{table*}

\subsection{Attribute value range (\texorpdfstring{$|S|$}{S})}

A second factor that defines the resilience of a device configuration to web fingerprinting is the \emph{cardinality}  of the reported attributes ($|S|$), defined as the number of distinct values an attribute can assume. If attribute $A$ can take 10 values whereas attribute $B$ can take 20 values, then attribute $B$ is (in principle) more useful to identify uniquely a user.
For instance, in desktop devices, if we factor out those attributes taking just 1 or 2 values (i.e., deterministic or binary attributes), Chrome offers a larger cardinality for all attributes compared to Safari and Firefox. As an example, an attribute with a high cardinality like the \emph{User-Agent} presents \mbR{(731, 72, 82)} potential values for (Chrome, Safari, Firefox), respectively.

However, although cardinality is crucial to determine an attribute's capacity to fingerprint individual users, we find examples where it might be partially misleading. For example, the attribute "\emph{Screen: pixel left}" can take potentially \mbR{(2074, 286, 101)} different values in (Chrome, Safari, Firefox) for desktop devices, respectively. In practice, \mbR{(77.3\%, 85.3\%, 56.7\%)} of the samples in our \emph{fingerprint-browsers} dataset take just a unique value (0) for this attribute in (Chrome, Safari, Firefox), respectively. 
Therefore, the discrimination power of an attribute to define unique fingerprints depends not only on its cardinality but also on the frequency of appearance of its potential values.

\subsection{Value frequency by attribute (\texorpdfstring{$h$}{h})}

As proposed by Eckersley \cite{eckersley2010}, we use the \emph{entropy} to measure the frequency with which different values of a given attribute appear. The higher the entropy of an attribute, the more distributed the use of different values of such an attribute is. 

We compute the Shannon entropy of the attributes considered in our study using the following expression:

\begin{equation}
    H(X) = - \sum_{i=1}^{n}P(x_{i})\cdot \log_{b} P(x_{i}) 
    \label{eq:entropy_equation}
\end{equation}

Where $H$ is the entropy, $P$ is the probability distribution of the attribute's values, and $n$ is the total possible values for the attribute. The Shannon entropy is measured in bits and thus it uses a logarithm base equal to 2 ($b$ = 2).

To compare the entropy of attributes with different cardinality, we use a normalized version of Shannon's entropy ($h$) dividing the equation~\ref{eq:entropy_equation} by ${\log_{2}(M)}$, where $M$ represents the number of values for each attribute.

\begin{equation}
    h(X) = \frac{H(X)} {\log_{2}(M)}
    \label{tab:normShannonEntropy_equation}
\end{equation}

${\log_{2}(M)}$ is the maximum possible diversity index where the entropy is maximum, and all values of an attribute are equally frequent. Hence, the resulting normalized entropy ranges between 1 (the attribute's largest discrimination power) and 0 (the attribute takes a unique value; thus, it is deterministic and useless for fingerprinting users).

However, the normalized entropy alone may again mislead us in identifying attributes with high discrimination power. For instance, an attribute with a normalized entropy close to 1 but a low cardinality, e.g., equal to 2, does not have a strong discrimination power. It just divides users into two groups. Therefore, the entropy alone is not sufficient to assess the discrimination power of an attribute.

\subsection{Summary of the analysis}

Our analysis above reveals that the discrimination power of an attribute is defined by three factors, it must: 1) be reported, 2) have high cardinality, and 3) have high entropy.

Table \ref{tab:featuresAttributes} provides a summary of these three factors ($|N|$, $|S|$, {$h$}) for each attribute in the most relevant mobile app and browser-based configurations considered in this paper. Note that for attributes concatenated within a meta-attribute, we report the values for the meta-attribute.

The results in the table offer two main insights. On the one hand, we can assess that the configurations reported as most vulnerable to web fingerprinting in Section \ref{sec:results} are characterized by a higher number of informed attributes with high discrimination power (high cardinality and normalized entropy). For instance, the \texttt{{Chrome, Windows}} configuration, identified as highly vulnerable, reports \mbR{35 out of the 35 attributes listed in Table \ref{tab:featuresAttributes}, with 8 of these attributes having a cardinality $\geq$ 299 and a normalized entropy $\geq$ 0.20. Contrary, the \texttt{{Firefox, Linux}} configuration reports 30 attributes, with just 1 exceeding the indicated cardinality and normalized entropy thresholds}. Likewise, it is evident that mobile configurations generally report fewer attributes than desktop configurations. Furthermore, attributes' cardinality and entropy tend to be lower in mobile configurations compared to desktop configurations.

On the other hand, Table \ref{tab:featuresAttributes} reveals those attributes with the most significant discrimination power (highlighted in bold). This represents very valuable information for browser and mobile apps developers, which can design solutions to mitigate the power discrimination of these attributes, reducing the vulnerability of the corresponding device configuration.


\vspace{1mm}
\noindent Remark. While this section focuses on static fingerprinting characteristics, Appendix~\ref{appendix:c} presents a longitudinal analysis of fingerprinting complexity over time, based on the joint entropy of full fingerprints across device configurations.

\section{Countering web fingerprinting}
\label{sec:solution}

The efficiency of fighting fingerprint is nowadays (and will be) an added value feature, which might positively affect the market share of a given browser. The browser industry is aware of this. As described in Section \ref{sec:background}, they have (or are) developed(ing) different techniques to counter fingerprinting. These techniques, in essence, try to block the reporting of specific attributes or to reduce their cardinality and/or entropy. Our results in Section \ref{sec:results} indicate that certain browsers already implement quite efficient solutions (e.g., Safari or Firefox) while others (e.g., Chrome) provide an almost negligible defense against web fingerprinting. 

The question we raise is whether there is room for improvement. To answer this question, we want to avoid complex solutions, which, although probably very efficient, require major changes in browsers and mobile apps and thus are very unlikely to be adopted by developers. Following this principle, we then reformulate our question to whether there are simple (yet efficient) approaches that would substantially improve the resilience to web fingerprinting offered by current browsers. We anticipate that the answer is yes.

\begin{figure}[t]
\centering
\includegraphics[width=\linewidth]{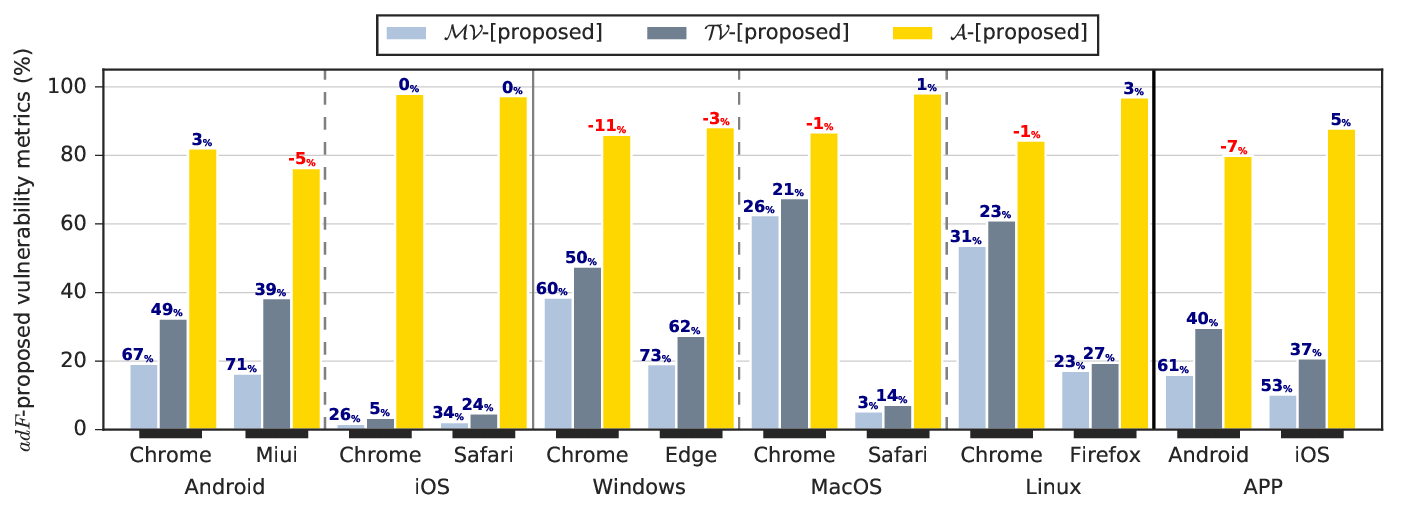}
\caption{\mbR{The effectiveness of the proposed anti-fingerprinting solution is indicated by $\mathcal{MV}$, $\mathcal{TV}$ and $\mathcal{A}$ for each device configuration. Values above bars denote positive (blue) or negative (red) percentage variation compared to metrics without \emph{ShieldF}.}}
\label{fig:benchking_adFP_metrics_browsers_apps}
\end{figure}

We propose \emph{ShieldF},  a solution which relies on the most common technique already implemented by several browsers,  this is, blocking/obfuscating the reporting of several attributes. This technique is simple, and browser developers are already using it, so it is likely that they are willing to adopt \emph{ShieldF}. The key difference between \emph{ShieldF} and existing solutions comes from the selection of the group of attributes to block. We propose to block those attributes which (1) have the largest discrimination power (i.e., highest cardinality and normalized entropy) and (2) do not affect (or have a very limited impact in) the end user experience\footnote{For instance, an attribute revealing the screen size cannot be blocked even if its discrimination power is high. The screen size is required to adapt the content of a web-page to the screen size so it renders adequately.}. To make an informed decision on the selected attributes to block, we rely on the results presented in Table \ref{tab:featuresAttributes}, which reports the discrimination power (cardinality and normalized entropy) of all considered attributes in our study. In particular, we propose to block those attributes with a cardinality $>$ \mbR{25} \footnote{Except for \emph{Bluetooth availability}, since it is characterized by a binary value.} and a normalized entropy $\geq$ \mbR{0.10}, which we are certain do not affect the browsing experience. The \mbR{12} selected attributes are highlighted in bold in Table \ref{tab:featuresAttributes}. Note this selection also implicitly accounts for correlated signals: we only consider an attribute mitigated when it cannot be inferred through other unblocked sources. Our proposal also applies to mobile apps with two exceptions: \emph{User Permissions state} and \emph{Bluetooth availability} attributes, because they are already unreported in mobile apps.

We assess the performance of \emph{ShieldF} in a threefold manner. First, we measure the reduction in the web fingerprinting vulnerability it achieves for all considered device configurations. Second, we compare its performance with the one offered by the anti-fingerprinting solutions implemented by popular browsers (Chrome, Safari, Firefox). Note that we do not consider other browsers for different reasons: they have a low market share (MIUI, Tor, Brave), they are implemented using a Chromium Engine (Edge) and thus are expected to offer a similar performance to Chrome, their anti-fingerprinting solution affects the users' browsing experience (Tor), or their ads do not allow fingerprinting from ads (Brave). Furthermore, we also compare \emph{ShieldF} performance with those solutions and techniques aimed at preventing unwanted web fingerprinting that adopt a similar methodology.

Note that for each specific solution considered in this section (those proposed by different browsers or ours), we re-train the Fingerprint Classifier from our \emph{adF} fingerprinting system. Hence the reported $\mathcal{MV}$ is the actual measurable vulnerability for each solution.

\subsection{Fingerprinting vulnerability reduction}

Figure \ref{fig:benchking_adFP_metrics_browsers_apps} shows $\mathcal{TV}$, $\mathcal{MV}$ and $\mathcal{A}$ for \emph{ShieldF} for all the considered browser-based (mobile app) configurations. Moreover, for each configuration, the number reported in the figure over the $\mathcal{TV}$ and $\mathcal{MV}$'s bar represents the reduction of the vulnerability (in percentage) compared to the original values of these metrics shown in Figure \ref{fig:browser_stats_adFP_metrics}.

\emph{ShieldF} reduces $\mathcal{TV}$ and $\mathcal{MV}$ in all considered configurations. In the case of $\mathcal{TV}$, such reduction ranges from \mbR{5\% \texttt{\{Chrome, iOS\}} to 62\% \texttt{\{Edge, Windows\}}}. 

As a complementary check, we also quantify the information carried by the blocked attribute set via joint entropy. We compute the joint entropy of the same pooled dataset before and after applying \emph{ShieldF} (formal definition in Appendix~\ref{appendix:c}). We observe a net drop of 2.15 bits (from 17.30 to 15.15), with an empirical $p$-value $<\!10^{-4}$ from a permutation test under the null of random assignment.



\subsection{Comparison to browsers' anti-fingerprinting proposals}

We compare the performance of \emph{ShieldF} with solutions currently implemented by the main browsers. We consider the \texttt{\{Desktop, Chrome, Windows\}} device configuration as the benchmark. We apply the most protective version of the solution (e.g., activating opt-in anti-fingerprint options) implemented by Safari, Firefox and Chrome introduced in Section \ref{sec:browsers_approaches} as well as our own solution to this configuration.

\begin{figure}[t]
\centering
\includegraphics[width=\columnwidth]{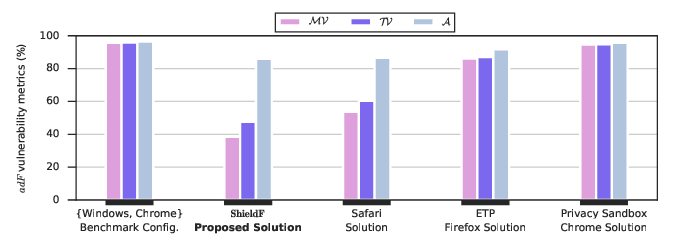}
\caption{\mbR{Comparing \emph{ShieldF} to Safari, Firefox, and Google solutions. We assess $\mathcal{MV}$, $\mathcal{TV}$, and $\mathcal{A}$ on a benchmark setup, \texttt{\{Desktop, Windows, Chrome\}}.}}
\label{fig:adFPmetrics_benchking}
\end{figure}

Figure \ref{fig:adFPmetrics_benchking} shows the values of $\mathcal{TV}$, $\mathcal{MV}$ and $\mathcal{A}$ for each solution. The results indicate that Google's privacy sandbox and Firefox's solution lead to limited improvements. Safari's solution would improve the resilience to web fingerprinting in the \texttt{\{Desktop, Chrome, Windows\}} configuration by a reduction of $\mathcal{MV}$ ($\mathcal{TV}$) about \mbR{44\% (37\%)}, which is still far from the \mbR{60\% (50\%)} improvement achieved by \emph{ShieldF}.

\subsection{Comparison with countermeasure}

While there is no definitive approach that can prevent web fingerprinting while maintaining the richness of a browser, the scientific community has developed techniques and solutions aimed at mitigating the effects of web fingerprinting.

Some works describe and analyze a specific method, such as Font fingerprinting \cite{nikiforakis2015privaricator}, Canvas fingerprinting \cite{acar2014web}, or WebGL fingerprinting \cite{wu2019rendered}. Others take a broader view and quantify the combined use of functions and APIs most commonly employed for web fingerprinting \cite{laperdrix2015mitigating}. 
While some solutions offer robust protection against browser fingerprinting, they often do so at the expense of usability \cite{noscript}. The greatest challenge researchers face is providing complete coverage of modified attributes, as even minor discrepancies can make users more visible to trackers \cite{boda2011user}. 

\begin{figure}[t]
\centering
\includegraphics[width=\columnwidth]{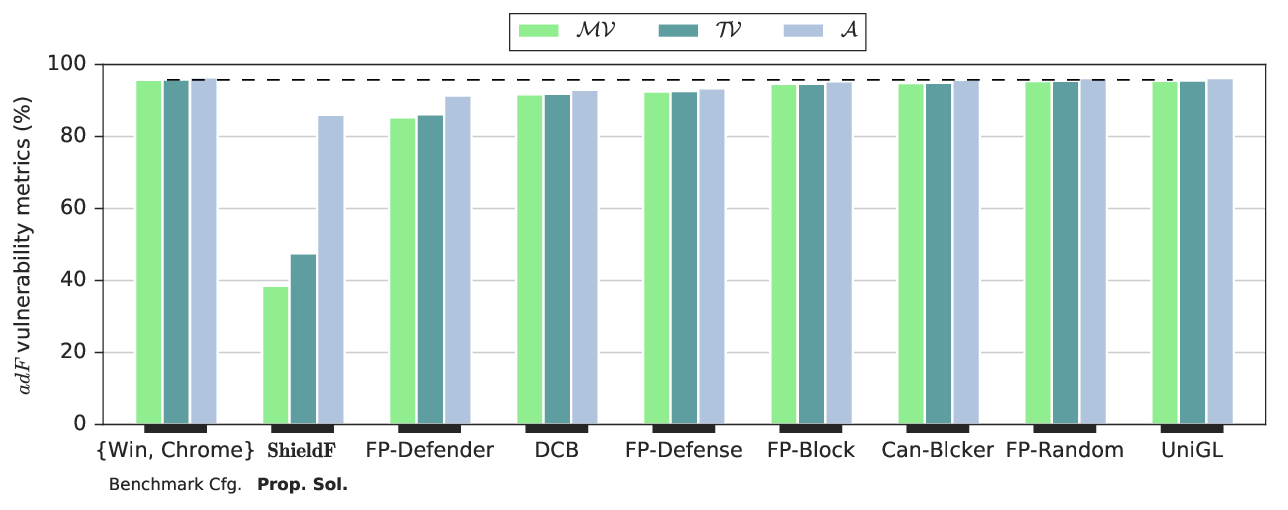}
\caption{\mbR{Comparing \emph{ShieldF} with techniques to mitigate web fingerprinting effects. We assess $\mathcal{MV}$, $\mathcal{TV}$, and $\mathcal{A}$ on a benchmark setup, \texttt{\{Desktop, Windows, Chrome\}}. Dashed line denotes $\mathcal{MV}$ of benchmark configuration \texttt{\{Desktop, Chrome, Windows\}}.}}
\label{fig:adFPmetrics_counterme}
\end{figure}

Figure \ref{fig:adFPmetrics_counterme} displays the values of $\mathcal{TV}$, $\mathcal{MV}$, and $\mathcal{A}$ for the  configuration \texttt{\{Desktop, Chrome, Windows\}} as a benchmark, as well as our solution, \emph{ShieldF}, for this configuration. 
The results indicate that in the general context of fingerprinting, solutions targeting a specific method like Canvas Blocker \cite{canvasbloquer_firefox}, UniGL \cite{wu2019rendered}, or even FP-Random \cite{laperdrix2017fprandom} —which focuses on defense techniques against Canvas fingerprinting, AudioContext fingerprinting, and unmasking browsers through JavaScript property ordering— offer protection that is seemingly marginal and makes almost no difference compared to not using them at all. 
Countermeasures like FP-Defense \cite{ajay2022defense}, FP-Block \cite{torres2015fp}, FP-Defender \cite{moad2021fingerprint}, and DCB \cite{baumann2016disguised}, which adopt a broader defense involving the combined use of JavaScript object-based functions and APIs, passive fingerprints, and HTTP headers, resulting in limited improvements in enhancing user privacy by preventing unwanted tracking. The FP-Defender solution would enhance resistance to web fingerprinting under the configuration of \texttt{\{Desktop, Chrome, Windows\}} by reducing $\mathcal{MV}$ ($\mathcal{TV}$) by approximately \mbR{11\% (10\%)}. 
However, this is still far from the improvement achieved by \emph{ShieldF}, which is \mbR{60\% (50\%)}.
On the other hand, the difference between $\mathcal{MV}$ and $\mathcal{TV}$ is larger with \emph{ShieldF} compared to other solutions. This implies a lower accuracy of the \emph{adF} fingerprinting system when our countermeasure \emph{ShieldF} acts, as it has fewer attributes with high cardinality and entropy. This means that our \emph{adF} fingerprinting system has a reduced capacity to identify unique fingerprints when \emph{ShieldF} is active compared to other countermeasures.

One of the main challenges surrounding web fingerprinting and, thus, the development of defensive techniques is the continuous evolution of current web browser technologies. A proposed countermeasure can become obsolete in a matter of months, as each new browser version that modifies, adds, or even removes an object or API has a potential impact on the web fingerprinting ecosystem. For instance, the removal of access to NPAPI plugins brought the end of an important source of information for web fingerprinting and, consequently, for mitigation efforts of some countermeasures \cite{boda2011user, torres2015fp, nikiforakis2015privaricator}. In contrast, the later introduction of features such as Canvas API \cite{acar2014web}, WebGL API \cite{moad2021fingerprint, wu2019rendered}, Battery Status API \cite{ajay2022defense}, and Web Audio API \cite{laperdrix2017fprandom}, revealed new vulnerabilities in web fingerprinting exploitation. As a result, the web fingerprints of the past are different from those we see today and will likely differ from those we will encounter in the coming years. Section 11 discusses assessed countermeasures in more detail.

\subsection{Empirical evaluation of \emph{ShieldF}}

We implement \emph{ShieldF} as an extension for any Chromium-based browser (Google Chrome, Microsoft Edge, Opera, Brave, etc).  First of all, we prove that the 12 selected attributes highlighted in Table \ref{tab:featuresAttributes} are effectively blocked by the \emph{ShieldF} extension. Moreover, \mbR{15} beta-testers have installed the \emph{ShieldF} extension and used it for at least \mbR{one month}. They have been instructed to report any issue arising from their browsing experience. At the time of submitting this paper, none of the beta-testers have reported any major issue.
The 15 beta-testers used desktop/laptop devices with up-to-date default installations and did not report using any additional privacy-preserving mechanisms beyond default browser settings. Their configurations were: \texttt{\{Chrome, Windows\}} 8/15 (53.3\%), \texttt{\{Chrome, macOS\}} 5/15 (33.3\%), and \texttt{\{Edge, Windows\}} 2/15 (13.3\%).
We acknowledge that this initial evaluation cannot be considered exhaustive due to the limited number of beta-testers, but it offers initial empirical evidence that \emph{ShieldF} has a negligible impact on users' browsing experience.
The \emph{ShieldF} extension is publicly available in the Chrome Web Store\footnote{\emph{ShieldF} is available at: \url{https://chromewebstore.google.com/detail/shieldf/hdpnjiibilpldnnccggcpgilnpbdciia?hl=es&authuser=0}} to allow any interested users to test it. 

\vspace{2mm}
-Summary: We conclude that a simple and easy-to-adopt solution by browsers and mobile app developers like \emph{ShieldF} may substantially improve resilience against web fingerprinting. Note that our solution is based on a simple heuristic selection of attributes without the will of it being optimal, and thus, its performance can be considered a lower-bound improvement. Each browser developer can take \emph{ShieldF} code  and results reported in Table \ref{tab:featuresAttributes} as a starting point to define its optimized selection of attributes to be blocked in its browser. 

Nevertheless, a crucial aspect of web fingerprinting and, consequently, mitigation solutions is the understanding that defense techniques quickly become deprecated. Therefore, this demands continuous monitoring of countermeasures to ensure they are regularly updated, guaranteeing that users remain protected against fingerprinting.

\begin{table*}[t]
\centering
\caption{\mbR{Overview of large-scale fingerprinting studies: uniqueness (\emph{unq}), no. collected attributes (\emph{No. att}) and $H_m$.}}
\label{tab:overviewFingerprinting}
\resizebox{\textwidth}{!}{
\begin{tabular}{l|c|cc|cc|cc|ccc|}
\cline{2-11}
\multicolumn{1}{c|}{\multirow{2}{*}{}} & Eckersley\cite{eckersley2010} & \multicolumn{2}{c|}{Laperdrix et al.\cite{laperdrix2016beauty}} & \multicolumn{2}{c|}{Gómez-Boix et al.\cite{gomezBoix2018hiding}} & \multicolumn{2}{c|}{Andriamilanto et al.\cite{andriamilanto2021largescale}} & \multicolumn{3}{c|}{\emph{adF}}            \\ \cline{2-11} 
\multicolumn{1}{c|}{}                  & desktop                                           & desktop                                   & mobile                                  & desktop                                   & mobile                                   & desktop                                        & mobile                                         & desktop      & mobile     & app     \\ \cline{1-1}
\multicolumn{1}{|l|}{unq}              & 94.20\%                                           & 89.40\%                                   & 81\%                                    & 35.70\%                                   & 18.50\%                                  & 84.40\%                                        & 39.90\%                                        & 85.61\%      & 49.93\%    & 36.80\% \\ \hline
\multicolumn{1}{|l|}{No. att}          & 6                                                 & \multicolumn{2}{c|}{17}                                                             & \multicolumn{2}{c|}{17}                                                              & \multicolumn{2}{c|}{262}                                                                        & \multicolumn{2}{c}{66}    & 35      \\ \hline
\multicolumn{1}{|l|}{$H_m$\footnotemark}            & 18.843                                            & \multicolumn{2}{c|}{16.859}                                                         & \multicolumn{2}{c|}{20.98}                                                           & \multicolumn{2}{c|}{21.983}                                                                     & \multicolumn{2}{c}{16.64} & 15.66   \\ \hline
\end{tabular}
}
\end{table*}
\footnotetext{$H_m$ is the maximum possible diversity index where the entropy is maximum, and all values of an attribute are equally frequent.}

\section{Related Work}
\label{sec:related_work}

\subsection{Fingerprinting on the web}

The profound privacy risks tied to web fingerprinting have motivated the research community to address the topic.
The existing literature has extensively examined the feasibility of employing various JavaScript APIs for fingerprinting devices such as Canvas \cite{acar2014web}, WebGL API \cite{cao2017cross}, available fonts \cite{gomezBoix2018hiding}, Web Audio API \cite{cao2017cross}, Battery Status API \cite{diaz2015leaking} or extensions \cite{laperdrix2021fingerprinting}, as well as combinations of attributes; Mayer \cite{mayer2009any} conducted, back in 2009, one of the pioneering works to uniquely identify individual devices. In particular, the author collected the content from \emph{navigator}, \emph{screen}, \emph{navigator.plugins}, and \emph{navigator.mimeTypes} attributes to compute devices fingerprints.
In addition, the research community has conducted different studies based on the measurement and web crawling techniques to assess to what extent web fingerprinting is used \cite{acar2014web}.
Most of the previous studies focus on desktop devices. However, the rapid growth of the mobile web and the development of its underlying technology has enabled the possibility to implement fingerprinting techniques to identify mobile devices uniquely \cite{tiwarik2023androidhybridapps}. Eubank et al. \cite{eubank2013shining} conducted one of the earliest studies on mobile web tracking, analyzing cookies and HTTP headers used by browsers and revealing distinctions between the fingerprints of mobile and desktop devices. 
A literature review can be found in the surveys by Laperdrix et al. \cite{laperdrix2020fpSurvey} and Zhang et al. \cite{zhang2022survey}).

In the context of our work, few papers have attempted large-scale analysis to assess users' exposure to web fingerprinting. A seminal work in the field by Eckersley \cite{eckersley2010} performed the first large-scale analysis of web fingerprinting. By leveraging the Panopticlick \cite{eckersley2010} website, the authors analyzed attributes collected from HTTP headers, JavaScript APIs, and Flash and Java plugins. 
More recently, Laperdrix et al. \cite{laperdrix2016beauty} performed the first large-scale analysis considering fingerprints from both mobile and desktop devices. They based their study on data collected from the AmIUnique website \cite{laperdrix2016beauty}.
These aforementioned works followed a similar approach, involving websites where volunteer users connect. These websites run scripts to compute the device's fingerprint and assess the uniqueness of each. 
Instead, Gómez-Boix et al. \cite{gomezBoix2018hiding} and Andriamilanto et al. \cite{andriamilanto2021largescale} collected fingerprints through a script deployed on one of the top 15 French websites, as per the Alexa traffic rank.
Similar to \cite{gomezBoix2018hiding, andriamilanto2021largescale}, Li et al. \cite{li2020wtmbf} deployed a fingerprinting tool on a real-world website and studied the effectiveness of web fingerprinting by collecting millions of samples.

We present a novel approach for measuring web fingerprinting using ads as measurement venues rather than websites. This offers some advantages such as: 1) ability to define targeted measurement experiments for specific device configurations, which allows for first time to reveal the vulnerability at the granularity of individual device configurations; 2) independence from the willingness of volunteers or website owners to provide data; 3) capacity to audit the vulnerability to web fingerprinting in mobile apps for the first time. 
Table \ref{tab:overviewFingerprinting} compares the major large-scale studies on web fingerprinting, including our own, across three dimensions: uniqueness, defined as the fraction of unique identified fingerprints for both desktop and mobile devices; number of collected attributes; and maximum diversity index. We observe that our methodology aligns well with findings of the most recent study \cite{andriamilanto2021largescale} in terms of uniqueness.

\subsection{Fingerprinting countermeasures}

The research community has been devoted to finding solutions to countermeasure the fingerprint.
Finding a definitive method or solution that avoids fingerprinting does not exist without compromising the user's browsing experience. A widespread idea among the research community is to approach this issue to seek collective immunity, hide the user in the larger set of users, and try to standardize their attributes by overriding, randomizing, spoofing, etc. 
We find in the literature two types of proposals to counter fingerprinting.

The first group proposes different types of manipulation of the reported attributes to third-parties~\cite{laperdrix2015mitigating, Al_FannahNasserMohammed2020Tltl}. The solutions proposed by browser developers and introduced in Section~\ref{sec:background} fall into this group. We also find proposals from the research community. For instance, Blink \cite{laperdrix2015mitigating} or Mimic \cite{azad2020taming} propose to modify the content of the fingerprint by spoofing the value of certain attributes, whereas Canvas Blocker \cite{canvasbloquer_firefox} blocks the whole Canvas. UniGL \cite{wu2019rendered} proposes to universalize the fingerprint by removing discrepancies for the WebGL API between devices. The approach we propose in this paper, \emph{ShieldF}, belongs to this group. Inspired by the work of Laperdrix et al. \cite{laperdrix2016beauty}, we do a thorough exploration of the attributes reported by browsers and mobile apps to third-parties to block those with a larger discrimination power, granted that they do not affect the end user's browsing experience. 

The second group aims at identifying and blocking fingerprinting scripts. For instance, FP-Inspector \cite{iqbal2021fingerprinting} uses machine learning classification algorithms and applies a layered set of restrictions to detect fingerprinting scripts. Privacy Badger \cite{privacybadger} prevents the execution of known fingerprinting scripts in the browser. Also, a broader extension such as NoScript \cite{noscript}, which blocks the execution of unwanted JavaScript scripts, can be applied in the context of fingerprinting. This group of solutions are complementary to the one presented in this paper.

\section{Conclusion}

In this paper, we present \emph{adF}, a novel system to assess the vulnerability of different device configurations to web fingerprinting. \emph{adF} leverages the widespread use of online ads as measurement venues for collecting device attributes from code embedded in ads. This gathered information is then processed to identify unique fingerprints.
We use \emph{adF} in several real ad campaigns, enabling us to estimate the current vulnerability of devices to web fingerprinting. 
The results obtained are concerning. In particular, \mbR{66\%} of desktop devices are vulnerable to web fingerprinting. In the case of mobile devices, \mbR{40\%} are vulnerable to web fingerprinting via browser-rendered ads, and \mbR{35\%} through mobile apps.

Furthermore, we conduct an in-depth analysis of the discrimination power associated with different attributes reported by browsers and mobile apps. These findings hold significant value for developers of browsers and mobile apps seeking to implement more effective countermeasures against web fingerprinting by blocking or controlling those attributes with a higher discrimination power.
Our detailed analysis also allows us to define a straightforward yet efficient solution to web fingerprinting. This solution proposes to block those attributes with the largest discrimination power, whose blocking does not affect the user experience in browsers or mobile apps. Our analysis shows that our proposal (easily adopted by browsers and mobile app developers) outperforms all anti-fingerprinting solutions offered by major browser (Chrome, Safari and Firefox), reducing device vulnerability to web fingerprinting up to \mbR{62\%}.
Given that collecting attributes from scripts embedded in ads may present more limitations compared to scripts embedded in web-pages' HTML code, we believe that results presented in this paper can be considered a lower bound of device vulnerability to traditional web fingerprinting.

\section*{Acknowledgement}
\label{sec:acks}

We thank anonymous reviewers for their constructive feedback.
This work is partially funded by framework of Recovery, Transformation and Resilience Plan funds, financed by European Union (Next Generation), under Grant \textit{Cátedras ENIA 2022 para la creación de cátedras universidad-empresa en IA} with reference ETD/1180/2022. 
This paper is also partially carried out within framework of Recovery, Transformation and Resilience Plan funds, financed by European Union (Next Generation). Through grant Análisis y miTIgación de riesgos de seguridad y privaCIdad asociados a la explotación de datos PersonAles en OTTs (ANTICIPA). 
Further funding is allocated to ENTRUDIT, grant TED2021-130118B-I00 funded by MCIN/AEI/10.13039/5011000011033 and NextGeneration EU/PRTR funds.

\bibliographystyle{IEEEtran}
\bibliography{reference}

\section*{Biography}
\vspace{-12mm}
\begin{IEEEbiography}[{\includegraphics[width=1in,height=1.25in,clip,keepaspectratio]{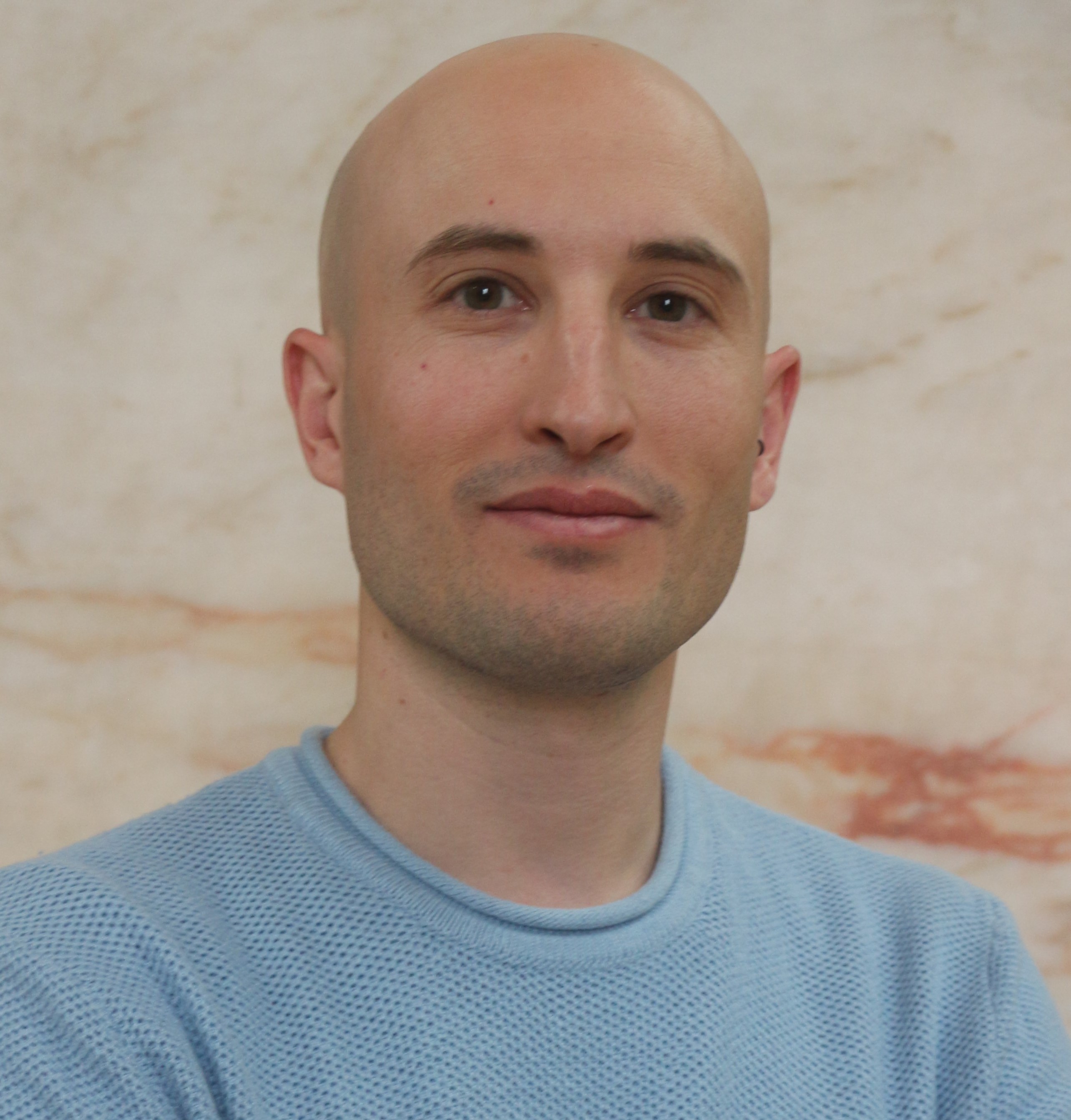}}]{Miguel A Bermejo-Agueda}
is a Ph.D. candidate in Telematics Engineering at Universidad Carlos III de Madrid. He received the B.Sc. in Telecommunications Engineering from Universidad Politécnica de Madrid and the M.Sc. in Data Science from the Universitat Oberta de Catalunya. His research interests include Internet measurements, online advertising, web sustainability, and privacy. He has published in venues such as ACM WWW and PoPETs, and has participated in EU H2020 projects.
\end{IEEEbiography}

\vspace{-15mm}
\begin{IEEEbiography}[{\includegraphics[width=1in,height=1.25in,clip,keepaspectratio]{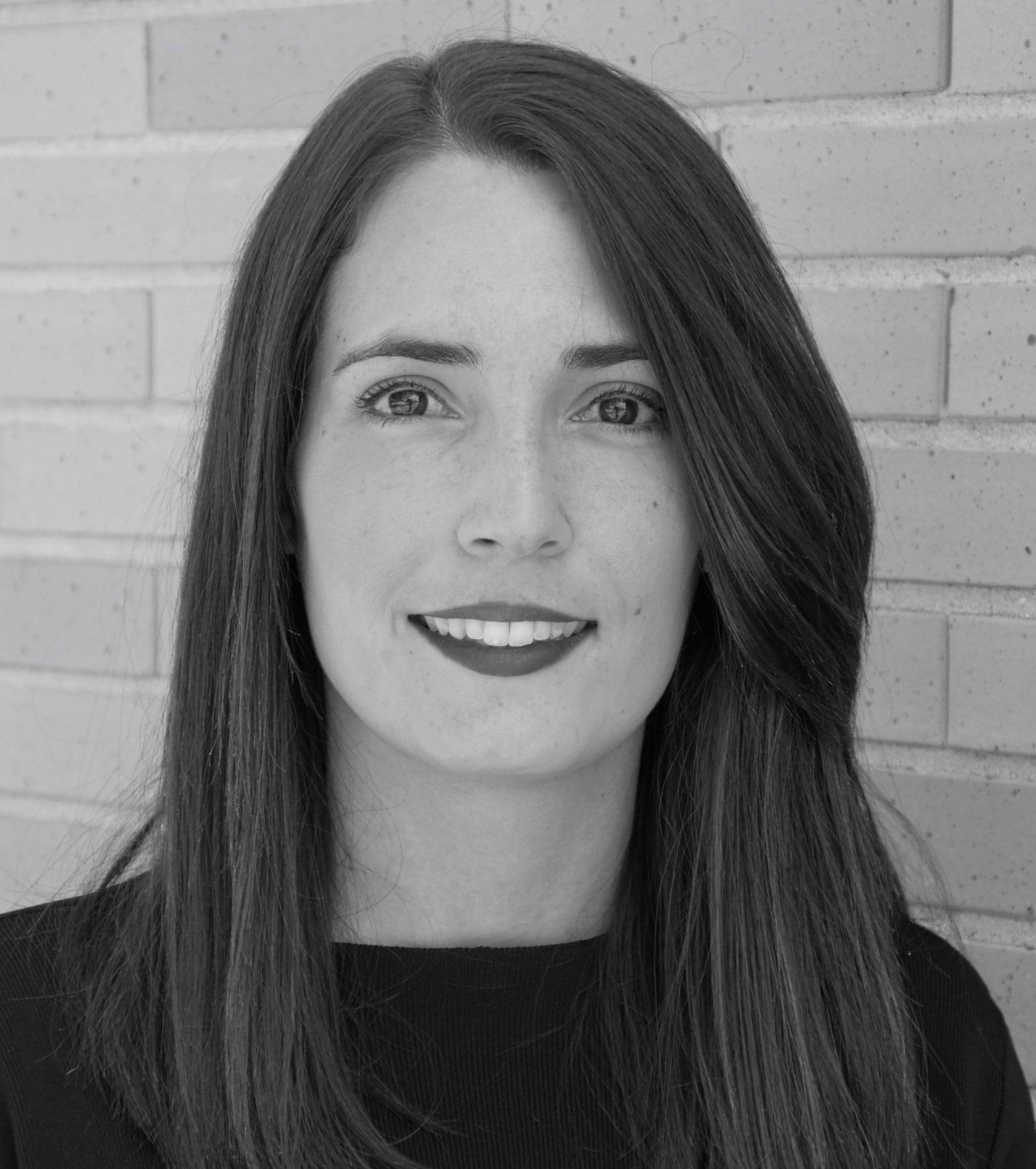}}]{Patricia Callejo}
is a Visiting Professor at Universidad Carlos III de Madrid, where she obtained her Ph.D. in Telematics Engineering (2020). She completed a research internship at ICSI, UC Berkeley, and received a RIPE Academic Cooperation Initiative (RACI) grant (RIPE 76, 2018). Her work spans Internet measurements, online advertising, and web transparency, with publications in ACM HotNets, ACM CoNEXT, and WWW, and participation in EU H2020 projects.
\end{IEEEbiography}

\vspace{-15mm}
\begin{IEEEbiography}[{\includegraphics[width=1in,height=1.25in,clip,keepaspectratio]{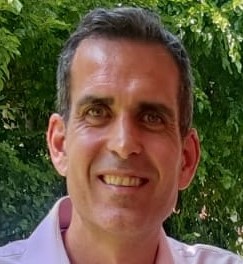}}]{Rubén Cuevas}
is an Associate Professor at Universidad Carlos III de Madrid, Deputy Director of the UC3M–Santander Big Data Institute and Courtesy Assistant Professor, University of Oregon (2012). He has coauthored 70+ papers in venues including PNAS, CACM, IEEE/ACM TON, WWW, USENIX Security, ACM CoNEXT, and CHI. His interests include online advertising, web transparency, personalization and privacy, online social networks, and Internet measurements.
\end{IEEEbiography}

\vspace{-15mm}
\begin{IEEEbiography}[{\includegraphics[width=1in,height=1.25in,clip,keepaspectratio]{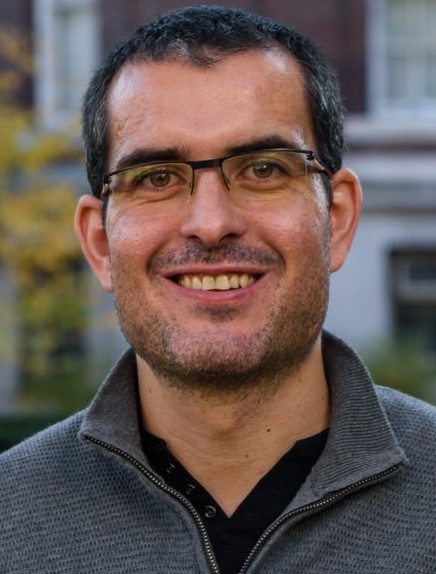}}]{Ángel Cuevas}
is an Associate Professor at Universidad Carlos III de Madrid. He has coauthored 70+ papers in top venues, including PNAS, IEEE/ACM Transactions on Networking (TON), WWW, USENIX Security, CHI, and Communications of the ACM (CACM). His research focuses on Internet measurements, web transparency, privacy, and P2P networks. He received the Best Paper Award at ACM MSWiM 2010 and the 2018 Emilio Aced Research Prize from the Spanish Data Protection Agency.
\end{IEEEbiography}

\vspace{-12mm}
\begin{IEEEbiography}[{\includegraphics[width=1in,height=1.25in,clip,keepaspectratio]{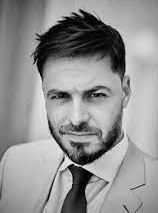}}]{Álvaro Mayol}
is Managing Director and Partner at TAPTAP Digital, previously serving as CPO and CTO. He is a Telecommunications Engineer from Universidad Politécnica de Madrid and the Illinois Institute of Technology. He began his career at Indra Systems (Electronic Warfare and Military Communications) and later led Solutions and Engineering at Geometry Global. In 2014 TAPTAP Digital acquired the company he co-founded, focused on proximity marketing and analytics.
\end{IEEEbiography}

\begin{appendices}
\section{Attributes considered by the \emph{adF} system}
\label{appendix:a}


Table \ref{tab:considered_attributes} provides detailed information of all the attributes initially considered to build the \emph{adF} system and the mapping from individual attributes to meta-attributes.
Low-level attributes that describe the same functional aspect (e.g., WebGL rendering parameters; navigator permissions) are concatenated into a single “meta-attribute”. While each constituent attribute may be only weakly discriminative on its own, jointly they capture a more stable, higher-level device property. Consolidating them removes redundancy and sparsity and yields a more compact representation—i.e., it effectively reduces dimensionality—while preserving the joint discriminative signal.

\begin{table*}
\centering
\captionsetup{labelformat=empty} 
\caption{Table IV: Attributes collected to build \emph{adF} system. With \texttt{asterisk} \texttt{(*)} attributes considered for fingerprinting in the mobile app ecosystem.}
\label{tab:considered_attributes}

\begin{minipage}[h][1\textheight][l]{0.5\textwidth}
\centering

\resizebox{0.88\textwidth}{!}{%
\begin{tabular}{|l|l|l|l|}
\hline
\multicolumn{1}{|c|}{\multirow{2}{*}{Meta-attributes}}                                  & \multicolumn{1}{c|}{\multirow{2}{*}{Attributes}} & \multicolumn{1}{c|}{\multirow{2}{*}{Reason}} & \multicolumn{1}{c|}{\multirow{2}{*}{Source}} \\
\multicolumn{1}{|c|}{}                                                                  & \multicolumn{1}{c|}{}                            & \multicolumn{1}{c|}{}                        & \multicolumn{1}{c|}{}                        \\ \hline
User-Agent                                                                              & \textbf{ua*}                                     & -                                            & JS object                                    \\ \hline
Time zone   offset                                                                      & \textbf{timeZoneOffset*}                          & -                                            & JS meth.                                     \\ \hline
\multirow{6}{*}{Bars settings}                                                          & menubar                                          & (quasi)-fixed value                          & \multirow{6}{*}{JS object}                   \\
                                                                                        & personalbar                                      & (quasi)-fixed value                          &                                              \\
                                                                                        & statusbar                                        & (quasi)-fixed value                          &                                              \\
                                                                                        & toolbar                                          & (quasi)-fixed value                          &                                              \\
                                                                                        & locationbar                                      & (quasi)-fixed value                          &                                              \\
                                                                                        & scrollbars                                       & (quasi)-fixed value                          &                                              \\ \hline
\multirow{5}{*}{Network Info}                                                           & downlink                                         & not stable over time                         & \multirow{5}{*}{JS API}                      \\
                                                                                        & effectiveType                                    & not stable over time                         &                                              \\
                                                                                        & rtt                                              & not stable over time                         &                                              \\
                                                                                        & saveData                                         & not stable over time                         &                                              \\
                                                                                        & type                                             & not stable over time                         &                                              \\ \hline
CPU cores                                                                               & \textbf{hw Concurrency*}                         & -                                            & JS object                                    \\ \hline
Device memory                                                                           & \textbf{device Memory*}                          & -                                            & JS object                                    \\ \hline
\multirow{8}{*}{Screen settings}                                                        & width                                            & correlated value                             & \multirow{8}{*}{JS object}                   \\
                                                                                        & height                                           & correlated value                             &                                              \\
                                                                                        & \textbf{colorDepth}                              & -                                            &                                              \\
                                                                                        & pixelDepth                                       & correlated value                             &                                              \\
                                                                                        & \textbf{orientation.angle*}                      & -                                            &                                              \\
                                                                                        & \textbf{orientation.type}                        & -                                            &                                              \\
                                                                                        & \textbf{screenLeft}                              & -                                            &                                              \\
                                                                                        & screenTop                                        & correlated value                             &                                              \\ \hline
\multirow{2}{*}{Nº touch points}                                                        & multitouch                                       & in maxTouchPoints                            & \multirow{2}{*}{JS object}                   \\
                                                                                        & \textbf{maxTouchPoints*}                         & -                                            &                                              \\ \hline
Content encoding                                                                        & content-encoding                                 & not stable over time                         & HTTP hdr                                     \\ \hline
Content type                                                                            & content-type                                     & not stable over time                         & HTTP hdr                                     \\ \hline
Referer                                                                                 & referer                                          & not stable over time                         & JS object                                    \\ \hline
URL                                                                                     & url                                              & not stable over time                         & JS object                                    \\ \hline
Platform                                                                                & platform                                         & in User Agent                                & JS object                                    \\ \hline
Languages                                                                               & \textbf{languages*}                              & -                                            & JS object                                    \\ \hline
Do Not Track                                                                            & doNotTrack                                       & -                                            & JS object                                    \\ \hline
Java enabled                                                                            & javaEnabled                                      & (quasi)-fixed value                          & JS object                                    \\ \hline
Web Driver                                                                              & webDriver                                        & (quasi)-fixed value                          & JS object                                    \\ \hline
PDF viewer   enabled                                                                    & \textbf{pdfViewerEnabled}                        & -                                            & JS object                                    \\ \hline
\multirow{5}{*}{Display window}                                                         & \textbf{availWidth*}                             & -                                            & \multirow{5}{*}{JS object}                   \\
                                                                                        & \textbf{availHeight*}                            & -                                            &                                              \\
                                                                                        & \textbf{availLeft}                               & -                                            &                                              \\
                                                                                        & \textbf{availTop}                                & -                                            &                                              \\
                                                                                        & \textbf{fullScreenEnabled*}                      & -                                            &                                              \\ \hline
\multirow{2}{*}{Storage}                                                                & usage                                            & not   stable over time                       & \multirow{2}{*}{AJAX API}                    \\
                                                                                        & \textbf{quota*}                                  & -                                            &                                              \\ \hline
navigator   object                                                                      & \textbf{window.navigator*}                       & -                                            & JS object                                    \\ \hline
List of   Plugins                                                                       & \textbf{plugins}                                 & -                                            & JS object                                    \\ \hline
\multirow{2}{*}{Cookies}                                                                & \textbf{cookieEnabled}                           & -                                            & \multirow{2}{*}{JS object}                   \\
                                                                                        & cookies                                          & not stable over time                         &                                              \\ \hline
List of MIME   type                                                                     & \textbf{mimeType}                                & -                                            & JS object                                    \\ \hline
\multirow{19}{*}{\begin{tabular}[c]{@{}l@{}}User Permissions\\      state\end{tabular}} & \textbf{accelerometer}                           & -                                            & \multirow{19}{*}{AJAX meth.}                 \\
                                                                                        & ambient-light-sensor                             & (quasi)-fixed value                          &                                              \\
                                                                                        & \textbf{background-fetch}                        & -                                            &                                              \\
                                                                                        & \textbf{background-sync}                         & -                                            &                                              \\
                                                                                        & \textbf{camera}                                  & -                                            &                                              \\
                                                                                        & \textbf{clipboard-read}                          & -                                            &                                              \\
                                                                                        & \textbf{clipboard-write}                         & -                                            &                                              \\
                                                                                        & \textbf{display-capture}                         & -                                            &                                              \\
                                                                                        & \textbf{geolocation}                             & -                                            &                                              \\
                                                                                        & \textbf{gyroscope}                               & -                                            &                                              \\
                                                                                        & \textbf{magnetometer}                            & -                                            &                                              \\
                                                                                        & \textbf{microphone}                              & -                                            &                                              \\
                                                                                        & \textbf{midi}                                    & -                                            &                                              \\
                                                                                        & \textbf{nfc}                                     & -                                            &                                              \\
                                                                                        & \textbf{notifications}                           & -                                            &                                              \\
                                                                                        & \textbf{payment-handler}                         & -                                            &                                              \\
                                                                                        & \textbf{persistent-storage}                      & -                                            &                                              \\
                                                                                        & push                                             & (quasi)-fixed value                          &                                              \\
                                                                                        & \textbf{screen-wake-lock}                        & -                                            &                                              \\ \hline
Media devices                                                                           & \textbf{mediaDevices*}                           & -                                            & AJAX meth.                                   \\ \hline
Canvas                                                                                  & \textbf{canvas*}                                 & -                                            & JS API                                       \\ \hline
List of Fonts                                                                           & \textbf{fonts*}                                  & -                                            & JS meth.                                     \\ \hline
Bluetooth                                                                               & \textbf{bluetoothAvailability}                   & -                                            & AJAX API                                     \\ \hline
\multirow{4}{*}{Battery status}                                                         & \textbf{charging}                                & -                                            & \multirow{4}{*}{AJAX API}                    \\
                                                                                        & level                                            & not stable over time                         &                                              \\
                                                                                        & chargingTime                                     & not stable over time                         &                                              \\
                                                                                        & dischargingTime                                  & not stable over time                         &                                              \\ \hline
\multirow{4}{*}{Audio context}                                                          & \textbf{baseLatency*}                            & -                                            & \multirow{4}{*}{JS API}                      \\
                                                                                        & \textbf{maxChannelCount}                         & -                                            &                                              \\
                                                                                        & \textbf{sampleRate*}                             & -                                            &                                              \\
                                                                                        & \textbf{state}                                   & -                                            &                                              \\ \hline
Frequency   analyser                                                                    & channelCount                                     & (quasi)-fixed value                          & JS API                                       \\ \hline
WebGL Vendor                                                                            & webglVendor                                      & in WebGL Renderer                            & JS API                                       \\ \hline
WebGL Renderer                                                                          & \textbf{webglRenderer*}                          & -                                            & JS API                                       \\ \hline
WebGL Extensions                                                                        & \textbf{webglExtensions*}                        & -                                            & JS API                                       \\ \hline
\end{tabular}%
}

\end{minipage}%
\begin{minipage}[h][1\textheight][l]{0.5\textwidth}
\centering

\resizebox{0.88\textwidth}{!}{%
\begin{tabular}{|l|l|l|l|}
\hline
\multicolumn{1}{|c|}{\multirow{2}{*}{Meta-attributes}}                                                                                                                                                     & \multicolumn{1}{c|}{\multirow{2}{*}{Attributes}}                                                          & \multicolumn{1}{c|}{\multirow{2}{*}{Reason}} & \multicolumn{1}{c|}{\multirow{2}{*}{Source}} \\
\multicolumn{1}{|c|}{}                                                                                                                                                                                     & \multicolumn{1}{c|}{}                                                                                                            & \multicolumn{1}{c|}{}                        & \multicolumn{1}{c|}{}                        \\ \hline
\multirow{3}{*}{\begin{tabular}[c]{@{}l@{}}WebGL\\      Attributes\end{tabular}}                                                                                                                           & antialias                                                                                                                   & (quasi)-fixed value                          & \multirow{3}{*}{JS API}                      \\
                                                                                                                                                                                                           & \textbf{powerPreference*}                                                                                                   & -                                            &                                              \\
                                                                                                                                                                                                           & desynchronized                                                                                                              & (quasi)-fixed value                          &                                              \\ \hline
\multirow{24}{*}{\begin{tabular}[c]{@{}l@{}}WebGL\\      Parameters\end{tabular}}                                                                                                                          & \textbf{\begin{tabular}[c]{@{}l@{}}ALIASED\_LINE-\\      \_WIDTH\_RANGE*\end{tabular}}                               & -                                            & \multirow{24}{*}{JS API}                     \\
                                                                                                                                                                                                           & \textbf{\begin{tabular}[c]{@{}l@{}}ALIASED\_POINT-\\      \_SIZE\_RANGE*\end{tabular}}                                    & -                                            &                                              \\
                                                                                                                                                                                                           & \begin{tabular}[c]{@{}l@{}}IMPLEMENTATION-\\      \_COLOR\_READ-\\      \_FORMAT\end{tabular}                              & (quasi)-fixed   value                        &                                              \\
                                                                                                                                                                                                           & \begin{tabular}[c]{@{}l@{}}IMPLEMENTATION-\\      \_COLOR\_READ-\\ \_TYPE\end{tabular}                                     & (quasi)-fixed   value                        &                                              \\
                                                                                                                                                                                                           & \textbf{\begin{tabular}[c]{@{}l@{}}MAX\_COMBINED-\\      \_TEXTURE\_IMAGE-\\      \_UNITS*\end{tabular}}                     & -                                            &                                              \\
                                                                                                                                                                                                           & \textbf{\begin{tabular}[c]{@{}l@{}}MAX\_CUBE\_MAP-\\      \_TEXTURE\_SIZE*\end{tabular}}                                      & -                                            &                                              \\
                                                                                                                                                                                                           & \textbf{\begin{tabular}[c]{@{}l@{}}MAX\_FRAGMENT-\\      \_UNIFORM-\\      \_VECTORS*\end{tabular}}                       & -                                            &                                              \\
                                                                                                                                                                                                           & \textbf{\begin{tabular}[c]{@{}l@{}}MAX-\\      \_RENDERBUFFER-\\      \_SIZE*\end{tabular}}                                  & -                                            &                                              \\
                                                                                                                                                                                                           & \begin{tabular}[c]{@{}l@{}}MAX\_TEXTURE-\\      \_IMAGE\_UNITS\end{tabular}                                              & (quasi)-fixed   value                        &                                              \\
                                                                                                                                                                                                           & \begin{tabular}[c]{@{}l@{}}MAX\_TEXTURE-\\ \_SIZE\end{tabular}                                                   & correlated   value                           &                                              \\
                                                                                                                                                                                                           & \textbf{\begin{tabular}[c]{@{}l@{}}MAX\_VARYING-\\      \_VECTORS*\end{tabular}}                                & -                                            &                                              \\
                                                                                                                                                                                                           & \textbf{\begin{tabular}[c]{@{}l@{}}MAX\_VERTEX-\\      \_ATTRIBS*\end{tabular}}                                & -                                            &                                              \\
                                                                                                                                                                                                           & \begin{tabular}[c]{@{}l@{}}MAX\_VERTEX-\\      \_TEXTURE-\\      \_IMAGE\_UNITS\end{tabular}                   & (quasi)-fixed   value                        &                                              \\
                                                                                                                                                                                                           & \textbf{\begin{tabular}[c]{@{}l@{}}MAX\_VERTEX-\\      \_UNIFORM-\\      \_VECTORS*\end{tabular}}                     & -                                            &                                              \\
                                                                                                                                                                                                           & \textbf{\begin{tabular}[c]{@{}l@{}}MAX\_VIEWPORT-\\ \_DIMS*\end{tabular}}                                & -                                            &                                              \\
                                                                                                                                                                                                           & RENDERER                                                                                                                  & in   WebGL Renderer                          &                                              \\
                                                                                                                                                                                                           & \textbf{SAMPLES*}                                                                                                            & -                                            &                                              \\
                                                                                                                                                                                                           & SAMPLE\_BUFFERS                                                                                                        & (quasi)-fixed   value                        &                                              \\
                                                                                                                                                                                                           & \begin{tabular}[c]{@{}l@{}}SHADING-\\ \_LANGUAGE-\\      \_VERSION\end{tabular}                                          & in   WebGL Renderer                          &                                              \\
                                                                                                                                                                                                           & STENCIL\_FUNC                                                                            & (quasi)-fixed   value                        &                                              \\
                                                                                                                                                                                                           & \textbf{\begin{tabular}[c]{@{}l@{}}STENCIL\_VALUE-\\      \_MASK*\end{tabular}}                                           & -                                            &                                              \\
                                                                                                                                                                                                           & \textbf{SUBPIXEL\_BITS*}                                                               & -                                            &                                              \\
                                                                                                                                                                                                           & VENDOR                                                                                                    & in   WebGL Renderer                          &                                              \\
                                                                                                                                                                                                           & VERSION                                                                                                                  & in   WebGL Renderer                          &                                              \\ \hline
\multirow{11}{*}{\begin{tabular}[c]{@{}l@{}}WebGL\\      ShaderPrecision:\\           .rangeMin\\           .rangeMax\\           .precision\\      -FRAGMENT (FR\_S)\\      -VERTEX (VR\_S)\end{tabular}} & FR\_S\_LOW\_FLOAT                                                                                                   & correlated   value                           & \multirow{11}{*}{JS API}                     \\
                                                                                                                                                                                                           & FR\_S\_MEDIUM\_FLOAT                                                                                      & correlated value                             &                                              \\
                                                                                                                                                                                                           & FR\_S\_HIGH\_FLOAT                                                                                       & (quasi)-fixed value                          &                                              \\
                                                                                                                                                                                                           & FR\_S\_LOW\_INT                                                                              & correlated value                             &                                              \\
                                                                                                                                                                                                           & FR\_S\_MEDIUM\_INT                                                                                           & correlated value                             &                                              \\
                                                                                                                                                                                                           & \textbf{\begin{tabular}[c]{@{}l@{}}FR\_S\_HIGH\_INT\\      {[}.rangeMax{]}*\end{tabular}}                & -                                            &                                              \\
                                                                                                                                                                                                           & VR\_SLOW\_FLOAT                                                                                                            & correlated value                             &                                              \\
                                                                                                                                                                                                           & VR\_SMEDIUM\_FLOAT                                                                                         & correlated value                             &                                              \\
                                                                                                                                                                                                           & VR\_SLOW\_INT                                                                                & correlated value                             &                                              \\
                                                                                                                                                                                                           & VR\_SMEDIUM\_INT                                                           & correlated value                             &                                              \\
                                                                                                                                                                                                           & VR\_SHIGH\_INT                                                                                                  & correlated value                             &                                              \\ \hline
\multirow{18}{*}{\begin{tabular}[c]{@{}l@{}}Audio formats\\           .supported\\           .smooth\\           .powerEfficient\end{tabular}}                                                             & \textbf{acc {[}.supported{]}*}                                                                                            & -                                            & \multirow{18}{*}{AJAX API}                   \\
                                                                                                                                                                                                           & x-wav                                                                                                                     & correlated value                             &                                              \\
                                                                                                                                                                                                           & mpeg                                                                                                          & correlated value                             &                                              \\
                                                                                                                                                                                                           & \textbf{aacp {[}.supported{]}}                                                                                & -                                            &                                              \\
                                                                                                                                                                                                           & wav                                                                                                             & correlated value                             &                                              \\
                                                                                                                                                                                                           & mp4-codec\_mp4a402                                                                                                & (quasi)-fixed value                          &                                              \\
                                                                                                                                                                                                           & mp4-codec\_ac-3                                                                & correlated value                             &                                              \\
                                                                                                                                                                                                           & mp4-codec\_ec-3                                                                                 & correlated value                             &                                              \\
                                                                                                                                                                                                           & mp4-codec\_alac                                                                                                 & (quasi)-fixed value                          &                                              \\
                                                                                                                                                                                                           & flac                                                                                                             & correlated value                             &                                              \\
                                                                                                                                                                                                           & mp4-codec\_flac                                                                                                & correlated value                             &                                              \\
                                                                                                                                                                                                           & ogg-codec\_flac                                                                                                  & correlated value                             &                                              \\
                                                                                                                                                                                                           & mp4-codec\_mp3                                                                                                   & correlated value                             &                                              \\
                                                                                                                                                                                                           & ogg-codec\_opus                                                                                                  & correlated value                             &                                              \\
                                                                                                                                                                                                           & webm-codec\_opus                                                                                                & correlated value                             &                                              \\
                                                                                                                                                                                                           & mp4-codec\_opus                                                                                                  & correlated value                             &                                              \\
                                                                                                                                                                                                           & ogg-codec\_vorbis                                                                                                 & correlated value                             &                                              \\
                                                                                                                                                                                                           & webm-codec\_vorbis                                                                                              & correlated value                             &                                              \\ \hline
\end{tabular}%
}
\end{minipage}
\end{table*}
\section{Out-Of-Distribution (OOD) evaluation}
\label{appendix:b}

To assess generalization beyond random, in-distribution validation, we evaluate the classifier under two out-of-distribution (OOD) settings: (i) \emph{cross-configuration} (leave-one-configuration-out, LOCO) and (ii) \emph{temporal} hold-outs across campaigns. In all cases, folds and splits are grouped by advertising ID (Ad-ID) so that all sessions from the same device remain on the same side of a split, preventing device-level leakage. The Ad-ID is used solely for ground-truth construction and grouping; it is never used as a model input.

Before any OOD experiment, we fix a single XGBoost configuration once, selected on a development split via 10-fold cross-validation grouped by Ad-ID. The very same hyperparameters are then reused for all OOD tests—no per-configuration re-tuning, no class reweighting, and no post-hoc threshold search—so the reported OOD figures constitute a conservative lower bound for a realistically pre-trained model.

To ensure reliability and avoid small-sample artefacts, we pre-specify inclusion thresholds independent of outcomes: (i) LOCO (cross-configuration): include \texttt{\{browser, OS\}} configurations with $\geq$\,2{,}000 unique Ad-IDs; (ii) Temporal hold-out: require $\geq$\,3 non-overlapping time windows with $\geq$\,3{,}000 unique Ad-IDs per window and disjoint Ad-IDs across time.

Due to configurations differ markedly in both class priors and test-set size, we report ROC-AUC (ranking, prior-agnostic) and balanced accuracy (thresholded, prior-sensitive), with 95\% nonparametric bootstrap confidence intervals. To summarize across highly unbalanced test sizes (from $\approx$2.3k to $\approx$63k Ad-IDs), we provide three aggregates: macro (equal weight per configuration), size-weighted (by Ad-IDs/rows), and micro-pooled (concatenated test predictions).

\paragraph*{Leave-One-Configuration-Out (LOCO)}
we hold out one \texttt{\{browser, OS\}} configuration for testing and train on the rest (no Ad-ID overlap). Eight high-support configurations meet the inclusion criterion. Table \ref{tab:ood_summary} shows aggregated results: macro AUC $0.729$, macro balanced accuracy $0.579$; micro-pooled over all held-out tests: AUC $0.706$ [0.703, 0.709], balanced accuracy $0.652$ [0.650, 0.654]. As expected under sharp prior shifts—e.g., \texttt{\{Chrome, Windows\}} $\approx$96\% positives vs. \texttt{\{Chrome, iOS\}} $\approx$4\%—AUC transfers more stably across configs, whereas thresholded metrics reflect deployment 

\begin{table}[H]
\centering
\captionsetup{labelformat=empty} 
\caption{Table V: OOD evaluation. Values are point estimates with 95\% bootstrap CIs where applicable.}
\label{tab:ood_summary}
\setlength{\tabcolsep}{1.1pt}
\renewcommand{\arraystretch}{1.25}
\resizebox{\linewidth}{!}{%
\begin{tabular}{lcc}
\toprule
Setting & ROC-AUC & Balanced accuracy \\
\midrule
LOCO (macro) & 0.729 & 0.579 \\
LOCO (micro-pooled) & 0.706 \;[0.703, 0.709] & 0.652 \;[0.650, 0.654] \\
Temporal hold-out (2023–09/10) & 0.769 \;[0.765, 0.773] & 0.637 \;[0.634, 0.640] \\
\bottomrule
\end{tabular}
}
\end{table}

\noindent priors (e.g., near-chance balanced accuracy in high-positive regimes despite AUC $>0.62$).
This is precisely the scenario where simple per-config calibration (e.g., Platt, isotonic) on the training portion aligns decision thresholds without changing the model.

\paragraph*{Temporal hold-out.}
We train on the first three predefined windows (02-2022, 05/06-2022, 05-2023) and test prospectively on 09/10-2023. With the same fixed hyperparameters, the model attains AUC $0.769$ [0.765, 0.773] and balanced accuracy $0.637$ [0.634, 0.640] on the held-out window (see Table \ref{tab:ood_summary}), indicating temporal generalization under natural \texttt{\{browser, OS\}} drift.

OOD results complement the in-distribution, grouped cross-validation reported in the main text and extend the reach of the ground-truth statistic ($\mathcal{TV}$) by providing deployable, identifier-free risk estimates on unseen configurations and future windows. Residual, prior-driven threshold mismatch is expected and can be addressed via light-weight calibration at deployment time; we intentionally keep the reported OOD figures uncalibrated to provide a conservative lower bound.

\section{Temporal Stability of Configuration-Level Web Fingerprinting}
\label{appendix:c}

To complement our core evaluation and address potential concerns regarding the temporal robustness of fingerprinting, we conduct a longitudinal analysis across four timeframes in our dataset: 02-2022, 05/06-2022, 05-2023, and 09/10-2023. Unlike prior works as Laperdrix et al. (2020) that examine the persistence of individual fingerprints over sessions, our approach focuses on the temporal evolution of the fingerprinting landscape at the configuration level, defined as the tuple \texttt{\{browser, OS\}}, as observed through repeated ad delivery campaigns. 

While Section~IX.C analyzes the individual attribute entropy, here we extend the perspective using the \textit{joint entropy} of all fingerprinting attributes to capture the overall structural complexity of the fingerprint space over time. Formally, the Shannon joint entropy of a set of attributes $X_1, X_2, \ldots, X_n$ is:

\vspace{-2mm}
\begin{multline}
H(X_1, X_2, ..., X_n) = - \sum_{(x_1, x_2, ..., x_n) \in \mathcal{X}} 
P(x_1, x_2, ..., x_n) \\
\times \log_2 P(x_1, x_2, ..., x_n)
\tag{5}
\end{multline}

where $\mathcal{X}$ is the set of all possible combinations of attribute values, and $P(x_1, x_2, ..., x_n)$ is the joint empirical probability of each such combination. This measure captures the richness and dispersion of fingerprints independently of class imbalance or sample size.

To ensure reliability, we only include configurations that (i) maintain at least 400 samples per timeframe, and (ii) exhibit a sample count variability (standard deviation over mean) below 70\% across timeframes. These constraints reduce bias from underrepresented or unstable datasets. The resulting set includes configurations for iOS and macOS, which meet these criteria consistently. Figure~\ref{fig:config_overtime_allBrowserAPPs} summarizes their temporal evolution.

\setcounter{figure}{5}
\begin{figure}[t]
\centering
\includegraphics[width=\linewidth]{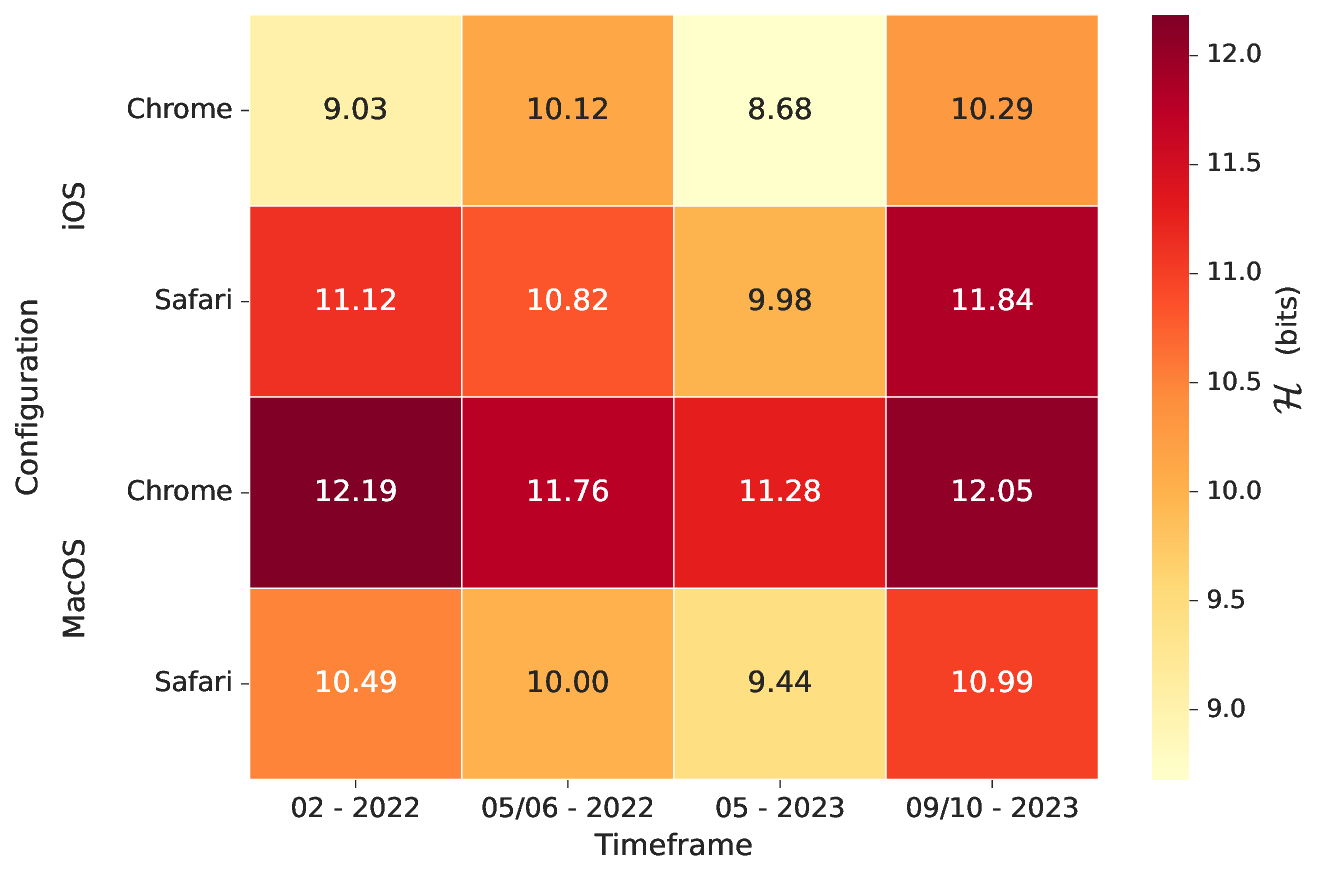}
\caption{Joint entropy ($\mathcal{H}$) over time for selected browser–OS configurations meeting sample stability criteria. Desktop setups show consistently higher entropy, while iOS-based configurations maintain lower and stable values.}
\label{fig:config_overtime_allBrowserAPPs}
\end{figure}

Among desktop setups, \texttt{\{Chrome, macOS\}} maintains high joint entropy (11.3–12.2 bits), indicating a persistent and diverse fingerprint space—suggesting little evolution in defenses. In contrast, \texttt{\{Safari, macOS\}} remains more stable and less expressive (10.0–11.0 bits), consistent with reduced API exposure.
On mobile, \texttt{\{Safari, iOS\}} shows a notably stable and low joint entropy (9.9–11.8 bits), reaffirming its well-known fingerprinting resistance. \texttt{\{Chrome, iOS\}} exhibits the lowest entropy values across all configurations ($<$10.3 bits), likely due to iOS sandboxing that limits Chrome’s attribute exposure. This behavior contrasts with desktop counterparts and supports previous findings that attribute diversity on iOS is intentionally constrained by platform-level controls.

No major non-monotonic shifts are observed among the selected configurations. Minor fluctuations (e.g., a dip for \texttt{\{Chrome, iOS\}} in 05 - 2023) correspond to lower sample sizes and should be interpreted with caution. These trends align with the findings of Laperdrix et al. (2020), who report that fingerprint changes are often driven by changes in browser APIs, software updates, and hardware context—not necessarily by user behaviour or population shifts. 

In summary, fingerprinting vulnerability does evolve over time, albeit unevenly across platforms and browsers. Joint entropy emerges as a robust metric to capture these temporal dynamics and validates the ability of \emph{adF} to operate as a longitudinal and configuration-aware fingerprinting auditor under real deployment conditions.

\section{Ethics considerations}
\label{appendix:d}
The research and experiments described in this paper obtained the IRB approval of our institution through the Data Protection Officer (DPO), a member of the Ethics Committee responsible for approving projects and experiments with potential data protection implications. 
Note that we take specific measures to prevent new risks for users (see Section V). In particular, we only consider fingerprints for those users for which we have an advertising ID. By doing so, we avoid creating any unique ID for users who do not have one. Moreover, we do not process any data from devices that have the do-not-track flag active in their browser.
Importantly, Advertising IDs are never used as features by our classifier. They are used \emph{exclusively} to (i) derive the ground-truth device identity for evaluation, and (ii) define non-leaking train/test splits via grouping.

Finally, our \emph{adTag} sends a total of 38.7kB per measurement to our back-end server. This is a negligible amount of data, which has a minimal impact on the overall data consumption of the device.

\end{appendices}

\end{document}